\documentclass[%
 reprint,
superscriptaddress,
 amsmath,amssymb,
 aps,
floatfix,
]{revtex4-2}

\usepackage{graphicx}
\usepackage{dcolumn}
\usepackage{bm}
\usepackage{hyperref}
\usepackage{placeins}
\usepackage[]{siunitx}
\usepackage{fixmath}
\usepackage{upgreek}
\usepackage[caption=false]{subfig}
\usepackage[nolist]{acronym}
\usepackage{nicefrac}
\usepackage{xcolor}
\usepackage[normalem]{ulem}

\newcommand{\vc}[1]{\boldsymbol{#1}}
\newcommand{\mt}[1]{\mathbf{#1}}

\newcommand{\pd}{\upPartial} 
\newcommand{\tpd}{\mathrm{d}}

\newcommand{\pdv}[1]{\upPartial_{#1}}

\newcommand{\E}{\mathrm{e}}

\newcommand{\del}{\updelta}

\newcommand{\linie}{\rule{2ex}{2pt}}

\definecolor{rot}{RGB}{174,96,96}
\definecolor{blau}{RGB}{90,123,163}
\definecolor{lila}{RGB}{134,86,160}
\definecolor{gruen}{RGB}{88,163,109}
\definecolor{grau}{RGB}{178,178,178}
\definecolor{lachs}{RGB}{255,94,94}
\definecolor{gelb}{RGB}{242,226,51}
\definecolor{mint}{RGB}{84, 179, 149}

\newsavebox{\foobox}
\newcommand{\slantbox}[2][0]{\mbox{%
    \sbox{\foobox}{#2}%
    \hskip\wd\foobox
    \pdfsave
    \pdfsetmatrix{1 0 #1 1}%
    \llap{\usebox{\foobox}}%
    \pdfrestore
}}
\newcommand\unslant[2][-.25]{\slantbox[#1]{$#2$}}
\newcommand{\upPartial}{\unslant\partial\kern-0.8pt}

\begin{document}


\title{The dynamics of a driven harmonic oscillator \\ coupled to pairwise interacting Ising spins in random fields}

\author{Paul Zech}
 \affiliation{Institute of Physics, Chemnitz University of Technology, 09107 Chemnitz, Germany.}%

\author{Andreas Otto}
 \affiliation{Institute of Physics, Chemnitz University of Technology, 09107 Chemnitz, Germany.}%
 \affiliation{Fraunhofer Institute for Machine Tools and Forming Technology, 09126 Chemnitz, Germany.}
 
\author{G{\"u}nter Radons}%
 \affiliation{Institute of Physics, Chemnitz University of Technology, 09107 Chemnitz, Germany.}%
 \affiliation{Institute of Mechatronics, 09126 Chemnitz, Germany.}




\date{\today}

\begin{abstract}
In general we are interested in dynamical systems coupled to complex hysteresis. Therefore as a first step we investigated recently the dynamics of a periodically driven damped harmonic oscillator coupled to independent Ising spins in a random field. Although such a system does not produce hysteresis, we showed how to characterize the dynamics of such a piecewise-smooth system, especially in the case of a large number of spins [P. Zech,  A. Otto, and  G. Radons, Phys. Rev. E 101, 042217 (2020)]. In this paper we extend our model to spin dimers, thus pairwise interacting spins. We show in which cases two interacting spins can show elementary hysteresis and we give a connection to the Preisach model, which allows us to consider an infinite number of spin-pairs. This thermodynamic limit leads us to a dynamical system with an additional hysteretic force in the form of a generalized play operator. By using methods from general chaos theory, piecewise-smooth system theory and statistics we investigate the chaotic behavior of the dynamical system for a few spins and also in case of a larger number of spins by calculating bifurcation diagrams, Lyapunov exponents, fractal dimensions and self-averaging properties. We find that the fractal dimensions and the magnetization are in general not self-averaging quantities. We show, how the dynamical properties of the piecewise-smooth system for a large number of spins differs from the system in its thermodynamic limit.
\end{abstract}

\maketitle


\section{Introduction} \label{sec:int}
Hysteresis phenomena can occur in many different fields, such as magnetic materials, elastics rubbers, liquid crystals or economy. The investigation of hysteretic systems is an ongoing field of research. For example, hysteresis plays a crucial role in the control of skyrmions via a strain-mediated magnetoelectric coupling \cite{Gusev2020}, in nonmonotonic field and temperature responses of magnetic superconductors \cite{Vlasko-Vlasov2020}, in the development of new high coercivity magnets \cite{Moorsom2020} or in evolutionary dynamics \cite{das2021}.

Many of the existing studies on hysteresis are related to the input-output behavior of hysteresis operators \cite{Bertotti2006, radons2008hysteresis, radons2008spectral1, radons2008spectral2, Jiang2010, schubert2017, urbanaviciute2018} or Ising like spin models \cite{Sethna1993, Lilly1996, Shukla2000, Sabhapandit2002, Deutsch2004}. On the other hand, hysteresis operators can be coupled to dynamical systems leading to a closed-loop dynamical system with hysteresis. In most of the works related to these systems, hysteresis is described by a bi-stable unit leading to a single elementary hysteresis loop in the input-output representation of the hysteresis operator. This kind of hysteresis is known as ``simple'' hysteresis. For example, there are some results on the conditions for global stability in systems with relay feedback \cite{Goncalves2001}, on multistability and hidden attractors in a DC/DC converter with hysteretic relay control \cite{ZHUSUBALIYEV2015} as well as results on systems with delayed relay control \cite{Sieber2006}. 
Also the dynamics of an oscillator with harmonic forcing and a single hysteretic relay has been studied in more detail \cite{Nagy2011,lelkes2021}.

A superposition of many elementary hysteresis loops or relays leads to the prominent \ac{PM} \cite{Preisach1935}, which is suitable for describing systems with ``complex'' hysteresis. ``Complex'' hysteresis is characterized by one major hysteresis loops and infinitely many inner subloops in the input-output representation \cite{Mayergoyz2003}. Examples of dynamical systems with ``complex'' hysteresis can be found in robot arm dynamics \cite{Ruderman2009}, in the ferroresonance phenomena in LCR circuits \cite{Lamba1997,Rezaei-Zare2007}, or in a mechanical system characterized by an oscillator with a ferromagnetic iron mass in an external magnetic field \cite{Donnagain2006,radons2013nonlinear}. 

A special case of ``complex'' hysteresis is given by an elementary play or backlash operator, which also appears in various dynamical systems \cite{kuhnen1999,Ruderman2019}. In the same manner as the \ac{PM} is defined by a superposition of elementary relay blocks, the so called Prandtl-Ishlinskii model for ``complex'' hysteresis is build of elementary play operators \cite{visintin2013}. There are various works on dynamical systems with this type of hysteresis, mainly in the context of control theory \cite{Krejci2001, Riccardi2014}.

Another prominent model, which shows ``complex'' hysteresis is the \ac{RFIM} at zero temperature. It can be used, for example, to model magnetic dipole moments of atoms. Originally the \ac{RFIM} was introduced to study phase transitions with a renormalization group approach \cite{Imry1975}. Later, also the hysteretic features of the \ac{RFIM} gained some interest \cite{Sethna1993, Perkovic1995, Shukla1996, Sethna2001}. 

In \cite{zech2020} we studied a degenerated case of a \ac{RFIM} without hysteresis, where nearest neighbor interactions were neglected. In this paper, as an extension of \cite{zech2020}, we consider an ensemble of independent spin pairs, which we will call ``dimers''.
Also quantum mechanical treatments of spin dimers can be found in the literature \cite{fowler1978,ramos2014}, but here we focus on a classical point of view. We analyse the system dynamics of such a gas of dimers at zero temperature and show, that hysteresis is possible. In this case two spins form an elementary hysteresis loop or relay and we show that this kind of hysteresis can be equivalently modeled by a \ac{PM}. Already for a single dimer, we can find chaotic solutions and bifurcations, which are typical for piecewise-smooth systems. We calculate the fractal dimensions of several chaotic attractors and compare the results of the many spin systems with the thermodynamical limit. This sheds some light on how the thermodynamic limit is approached by the piecewise-smooth system. We also determine the self-averaging properties of the fractal dimensions of the chaotic attractors.

In doing so, we also show, that the thermodynamical limit leads to a dynamical system with a nonlinearity in form of a generalized play operator \cite{Visintin1994,krasnosel1989,HASSANI2014}, known as the building block of the generalized Prandtl-Ishlinskii model \cite{aljanaideh2008}.

The organization of the paper is as follows. In Sec.\,\ref{sec:mod} we introduce the model and the different types of spin-spin interactions. Sec.\,\ref{sec:method} is about the methods, which are necessary for the analysis of the dynamical system with hysteresis. In particular, we describe specific methods for handling the discontinuities in piecewise-smooth systems, we derive a relationship between the spin dimers and the \ac{PM}, which is used to describe the system dynamics in the thermodynamical limit, and some remarks on the numerics and the calculation of the Lyapunov exponents are given. In Sec.\,\ref{sec:results} the results are presented, at first for the single dimer dynamics and later for many dimers, followed by a conclusion in Sec.\,\ref{sec:con} and an appendix.

\section{Model} \label{sec:mod}
 We study a periodically driven damped harmonic oscillator, which gets feedback from an ensemble of independent spin dimers at zero temperature. On the one hand, the driven harmonic oscillator is a classical and well-studied linear dynamical system. On the other, the spin dimers act as a model for complex hysteresis. In this section we briefly describe the two building blocks of the system and the coupling to each other. A more detailed derivation can be found in \cite{zech2020}.

\subsection{Oscillator model}
\label{sec:osci}

We consider the motion of a magnetizable point mass (which we call ``iron'' for simplicity) in an external magnetic field. The position $q$ of the iron mass is determined by the dynamics of a periodically driven harmonic oscillator in this field. In general the magnetization $M$ of the iron mass in dependency of the magnetic field $B$ is determined by the orientation of the intrinsic magnetic domains, which can be modeled e.\,g. by a \ac{RFIM}. Here, as a simplification we want to consider the magnetic system as an ensemble of independent spin dimers (a spin gas at zero temperature), also we assume a constant magnetic field gradient $B(q) \sim q$, such that the magnetization $M$ changes with the position $q(t)$ of the oscillator (cf. \cite{radons2013nonlinear}).

In dimensionless variables the equation of motion can be written as
\begin{equation}
     \ddot{q}(t) + 2 \zeta \dot{q}(t) + q(t) = \cos \Omega t + F_\text{hys}[q](t) \label{eq:eom_1},
\end{equation}
where $q(t)$ denotes the position of the oscillator at time $ t $, $ \zeta $ is the damping ratio of the system, and $ \Omega $ is the scaled excitation frequency. The mass of the oscillator is normalized to one and time $t$ is rescaled such that the eigenfrequency of the oscillator is equal to one. The hysteresis force $ F_\text{hys}[q](t) $ is the force due to the magnetization of the of the iron mass, which is determined by the spin dimers, the local random fields and the external magnetic field, which changes linearly with the oscillator position $q$. Here, we assume that the hysteresis force depends linearly on the magnetization, i.\,e., $ F_\text{hys}[q](t) = C M[q](t) $, where $ C $ is the coupling constant and $ M[q](t) $ denotes a functional that can depend on the whole trajectory of $ q(t') $ up to time $t$. Correspondingly $F_\text{hys}[q](t)$ is in general also a functional of the trajectory up to time $t$. This is due to the fact, that in contrast to \cite{zech2020} here the spin-spin interactions are not neglected, thus there is more than one metastable state, which the spin system can reach. Hence hysteresis is possible as described in Sec.\,\ref{sec:rfim} and Sec.\,\ref{sec:twospins}. Introducing the state variable $ \vc{x} = (q,v=\dot{q},\phi=\Omega t)^T $ in general, the oscillator dynamics can be described by
\begin{equation}
\dot{\vc{x}}(t)=
     \begin{pmatrix}
                v(t) \\
                 - 2\zeta v(t) - q(t) + \cos \phi(t) + C M[q](t) \\
                \Omega
                \end{pmatrix}  \label{eq:eom_2},
\end{equation}
where the ensemble of spin dimers is used to update the magnetization $M$ after each variation of $q(t)$ .

\subsection{Independent spin dimers at zero temperature}
\label{sec:rfim}

As mentioned before the spin system of the iron mass is modeled by an ensemble of independent spin dimers. Thus for an even number of $ N $ spins, we have $ N/2 $ spin dimers. A spin dimer is given by the two spins $\sigma_i$ and $ \sigma_{i-1} $ with even $ i=2,4,6,\ldots,N$, and as usual the magnetization per spin is given by the mean over all spins $ M = \frac{1}{N} \sum_{i=1}^N \sigma_i $. Nearest neighbor interactions, with the coupling strength $J$, only occur between the two spins of one spin dimer. So this system is not a spatially extended system. Also the state $ \sigma_i \in \{-1,+1\} $ of each spin is affected by the local field $ b_i $ and an external magnetic field proportional to $ q $, so in dimensionless variables the Hamiltonian can be written as:
\begin{equation}
    H = \sum_{k=1}^{N/2} \Bigl[ - J \sigma_{2k-1} \sigma_{2k} - \Bigl( \sigma_{2k-1}(q + b_{2k-1}) + \sigma_{2k} (q + b_{2k}) \Bigr) \Bigr] .
\end{equation}

Since in this paper we deal with spin dimers at zero temperature, the internal spin dynamics of the system is totally deterministic and can be described by the usual single spin flip dynamics \cite{vives2005,salvat2009}. This means a metastable state at time $ t $ is given, if each spin points in the direction of its local field $ F_i(t) $, that is
\begin{align}
    \sigma_i(t) &= \operatorname{sgn}(F_i(t)), \quad \forall i = 1,2,3,\dots,N  \label{eq:meta_cond} \\
    \text{with } F_i(t) &= \begin{cases}
                            J \sigma_{i-1}(t) + q(t) + b_i, \quad \text{$i$ even} \\ 
                            J \sigma_{i+1}(t) + q(t) + b_i, \quad \text{$i$ odd} \\ 
                            \end{cases} .                   
\end{align} 

Eq.\,\eqref{eq:meta_cond} is often called metastability condition. Here the spin system is first updated until a metastable state is reached and after that the oscillator position $q$ can change according to the new magnetization $M$ (see Sec.\,\ref{sec:numerics}). This property can be characterized as adiabatic limit \cite{Dahmen1996}. 

\subsection{Two pairwise interacting spins}
\label{sec:twospins}

\begin{figure}
    \centering
    \includegraphics[width=0.80\columnwidth]{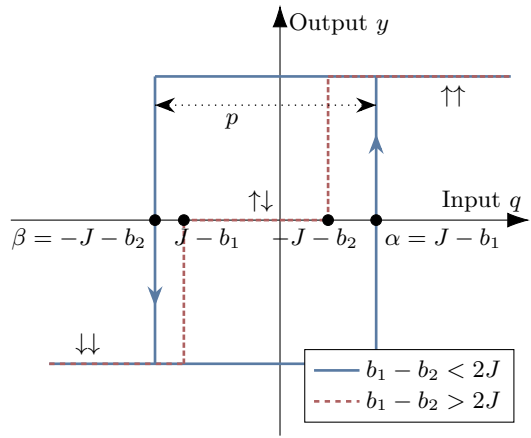}
   \caption{Graphical illustration of the magnetization (output $y$) of a spin dimer with $ J > 0 $ and $ b_{1} > b_{2} $. For $ 2J > b_1 - b_2 $ the system shows elementary hysteresis and therefore memory. In contrast, for $ 2J < b_1 - b_2 $ a different scenario occurs. Starting with both spins down and increasing the input $q$, at first one spin flips upward at $q=J-b_1$ and later at $q=-J-b_2$ the second spin flips upward.}
    \label{fig:elem_hys}
\end{figure}

We start by discussing the special case of $ N = 2 $ spins coupled to each other, i.\,e. with one spin dimer. In this case the metastability condition from Eq.\,\eqref{eq:meta_cond} reads
\begin{equation}
\label{eq:twospincond}
    \sigma_{1/2} = \operatorname{sgn}(J \sigma_{2/1} + q + b_{1/2}) ,
\end{equation}
where $ \sigma_{1/2} $ denotes either the ``first'' or the ``second'' spin of each spin dimer. Without loss of generality we assume $ b_{1} > b_{2} $. Then, from $ 2^N = 4 $ (not necessarily metastable) internal states, only three metastable states remain. The conditions to find the two spins in one of the three states can be derived from Eq.\,\eqref{eq:twospincond}. They are summarized in Table\,\ref{tab:cond} and illustrated in Fig.\,\ref{fig:elem_hys}. We can see, that qualitatively three different scenarios under a variation of the position $q$ are possible.
\begin{table}[!ht]
    \centering
    \begin{tabular}{|c|c|} \hline
        $\downarrow \downarrow$ & $q < J - b_{1}$ \\
        $ \downarrow \uparrow $ & not possible \\
        $\uparrow \downarrow$ & $J-b_1 < q < -J-b_2 $ \\
        $\uparrow \uparrow$ & $-J - b_{2} < q$ \\ \hline
    \end{tabular}
    \caption{Possible metastable states and corresponding conditions for two coupled spins with $ J > 0 $. Since we assume $ b_{1} > b_{2} $, the second configuration is not a metastable state. The third configuration is only possible, if $ J-b_1 < -J-b_2 $, i.\,e. for $ 2J < b_1-b_2 $.}
    \label{tab:cond}
\end{table}
The first scenario occurs for $b_{1} - b_{2}>2J$. In this case, for $q<J-b_1$ the two spins are pointing downwards ($\sigma_1=\sigma_2=-1$). By increasing $q$, the first spin $ \sigma_1 $ flips up at $ q = J - b_{1} $ and the second spin $ \sigma_2 $ flips at $ q = -J-b_{2} $ (red dashed line in Fig.\,\ref{fig:elem_hys}). In the second scenario, for $b_{1} - b_{2}=2J$, only one jump at $q=J-b_1=-J-b_2$ occurs, where both spins flip at the same position. In the third scenario for $b_{1} - b_{2}<2J$, both spins pointing downwards for $q<-J-b_2$. By increasing $q$, both spins flip upward at $ q = \alpha = J - b_{1} $. However, starting at $ q > J - b_{1} $, where both spins are in the upper state, and decreasing $q$, both spins flip downward at $ q = \beta = -J-b_{2} $ with $\beta<\alpha$ (blue solid curve in Fig.\,\ref{fig:elem_hys}). Hence, in this case, for $\beta< q < \alpha$ two metastable states exist, where $\beta$ and $\alpha$ are the lower and higher jump value, respectively. This means that in scenario three an elementary hysteresis loop occurs for two interacting spins, and the internal state depends on the history of $q$, or in other words, the system shows memory.  

\begin{figure}
    \centering
    \includegraphics[width=0.95\columnwidth]{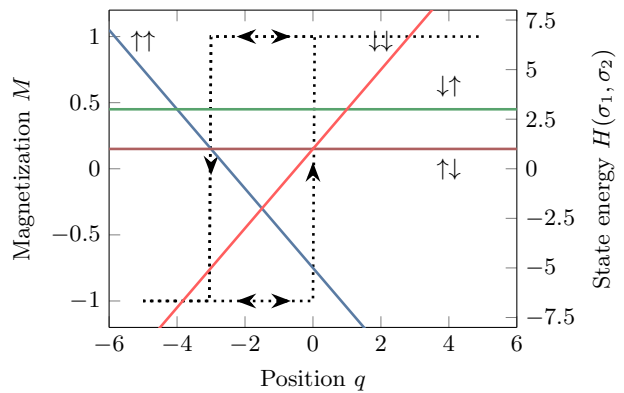}
    \caption{Numerical validation of the behavior of two spins with the coupling constant $ J = 2 $ and the local fields $ b_1 = 2 $, $ b_2 = 1 $. The straight lines show the energies $H(\sigma_1,\sigma_2)$ of the different internal states of the system. It can be seen, that for a single spin flip dynamics only three states of the system can be reached ($\downarrow \downarrow$,$\uparrow \downarrow$,$\uparrow \uparrow$), whereas one state is impossible to reach ($\downarrow \uparrow$). Also the related threshold values $ \beta = -3 $ and $ \alpha = 0 $ are given by the change of the energetically favorable states, indicated by the corresponding intersection points.}
    \label{fig:elem_hys_ex}
\end{figure}

A numerical example with $ J=2 $, $ b_{1} = 2 $ and $ b_{2} = 1 $ is presented in Fig.\,\ref{fig:elem_hys_ex}. One can see the output, the magnetization $M$ (dotted black line) depending on its input $ q $. Indeed, the first spin flips upward at $q=J-b_1=0$, and the second spin flips downward at $q=-J-b_2=-3$, which are the boundaries $\alpha$ and $\beta$ of the elementary hysteresis loop, respectively. Thus, the width $p$ of the loop is equal to $p=2J-b_1+b_2=3$. In addition, we have plotted the energy $H(\sigma_1,\sigma_2)$ of the four spin configurations $\{\sigma_1,\sigma_2\}$, $\sigma_i=\pm 1$ as functions of q, which can explain the spin flips from an energetic point of view. For $q<\beta=-3$ all spins are pointing down $ \downarrow \downarrow $, which is the state with the lowest energy. By increasing $q$, the energy level of the downward state increases, and at $ q = -1.5 $, the $ \uparrow \uparrow $-state becomes the state with the lowest energy. However, this state cannot be reached by a single spin flip (cf. Sec.\,\ref{sec:rfim}), and therefore, the system stays in the metastable state $ \downarrow \downarrow $. Later, at $ q = \alpha=0 $ the $ \uparrow \downarrow $-state has the same energy as the $ \downarrow \downarrow $-state, as a consequence, the first spin flips, and the resulting state is $ \uparrow \downarrow $. However, this state is not metastable because a second single spin flip is possible, for which the system can reach the energetically favorable $ \uparrow \uparrow $-state. Since the spin system is only updated if a metastable state is reached, at $ q = \alpha = 0 $ the system immediately jumps form $ \downarrow \downarrow $ to $ \uparrow \uparrow $. A similar procedure occurs for decreasing values of $ q $, where the reverse jump occurs at $q = \beta = -3$. From this energetic picture it is also clear, that only three of the four possible states can be reached, because the $ \downarrow \uparrow $-state has always a larger energy than the $ \uparrow \downarrow $-state. These results are fully consistent with the results in Table~\ref{tab:cond}. 

\subsection{Many spin dimers}
\label{sec:manyspindimers}
In this paper we study the special case of spin dimers, where each of the dimers has the same width of the hysteresis loop, therefore without loss of generality and consistent with the case of a single spin dimer, we assume $ b_{i-1} > b_i$, with $ i=2,4,\ldots,N$. Now we choose the two local field values $ b_i$ and $ b_{i-1}$ such that an elementary hysteresis loop is formed and that each loop has the same width ($p=2J-b_{i-1}+b_i>0$). Since by definition $b_{i-1} > b_i$, we obtain $ 0 < p < 2J $ for possible fixed widths $p$ of the elementary hysteresis loops. The higher and the lower jump values are given by  $\alpha_k=J-b_{2k-1}$ and $\beta_k=-J-b_{2k}$, respectively, where $k=1,\dots,N/2$ numbers the spin dimers. 

This paper deals, on one hand, with spins in a random field. On the other hand the hysteresis loops associated with the spin dimers should all have the same width $p$. Therefore, if one local field value, for example $ b_{i-1}$, is chosen randomly, the other field value $ b_i$ is automatically determined by $ b_i = p - 2J + b_{i-1}$. Correspondingly the jump values are given by 
\begin{align}
  \alpha_k &= J-b_{2k-1} \\
  \beta_k &= J-b_{2k-1}-p . \label{eq:jump_val}
\end{align}
For practical reasons instead of choosing one of the field values $ b_{i-1} $ or $ b_{i} $ for each dimer, we will choose the center $ s_k = (\alpha_k + \beta_k) / 2 $ of each elementary hysteresis loop to be Gaussian distributed and uncorrelated with $ \overline{s_k s_l} = R^2 \del_{kl} $, $ \overline{s_i} = 0 $ and determine the related values as $ b_{2k-1} = - s_k + J - \frac{p}{2} $ and $ b_{2k} = - s_k - J + \frac{p}{2} $. Here $\overline{X}$ denotes the average of $X$ over all realizations of the quenched disorder field $\{b_i\}$.   

\section{Method} \label{sec:method}
  \subsection{Piecewise-smooth system}
\label{sec:pws}

It is clear, that the coupled system contains continuous degrees of freedom of the harmonic oscillator as well as discrete degrees of freedom of the spins, and therefore, it can be treated by methods from piecewise-smooth system theory \cite{bernardo2008}. The investigation of such systems can be fields such as relay feedback systems \cite{Goncalvez2001}, gear dynamics \cite{theodossiades2000} or systems with dry friction \cite{galvanetto2001}, but also in the field of tapping-mode atomic force microscopy \cite{Zhao2005}. Here, we briefly introduce the theory of piecewise-smooth dynamical systems, that we have used to analyze our system. More details can be found in \cite{zech2020}.

In a system with an even number of $ N $ spins, we have $N/2$ spin dimers and both spins in one dimer only can be in the down state or in the up state. This gives us the state space $\{ \, q,v,\phi \, \mbox{mod}\,2\pi \, \} \times \{ \, \pm 1 \, \}^{N}$ , where the corresponding dynamics of the system takes place in a $\{ \, q,v,\phi \, \mbox{mod}\,2\pi \, \} \times \{ \, \pm 1 \, \}^{N/2}$-subspace only. In the following we call the discrete spin states \textit{sheets} $S_n$ with $ n = 1,\dots,2^{N/2} $. From the discussion above we know that the switching of the spin dimers depends only on the actual position $q$ of the system. Thus, the transition from one sheet $S_i$ to another sheet $S_j$ is given by the two-dimensional manifold $ q = q^*_{ij} $. Since different $S_n$ can show the same magnetization, there are only $N/2+1$ different values for the magnetization $ M_k = \frac{4k}{N} - 1 $, $k=0,\dots,N/2$. Hence, in case the system propagates within the smooth regions between the switching, Eq.\,\eqref{eq:eom_2} becomes
\begin{equation}
\label{eq:eom_4}
\dot{\vc{x}}(t)=
\begin{pmatrix}
v(t) \\
- 2 \zeta v(t)-q(t) + \cos \phi(t) + C M_k \\
\Omega
\end{pmatrix} .
\end{equation}

We will discuss this in more detail by looking at the two examples of two and four spins corresponding to one and two spin dimers. 

\begin{figure}
    \centering
    \includegraphics[width=1.0\columnwidth]{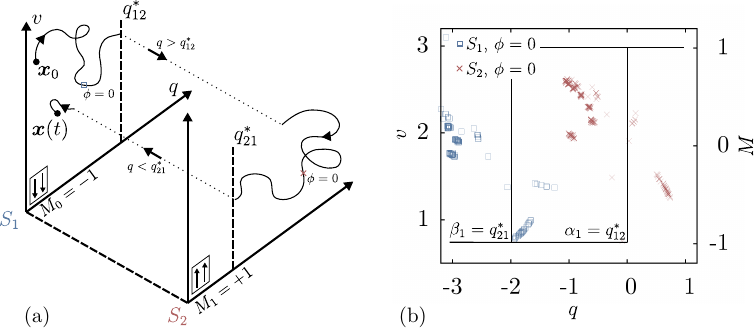}
    \caption{(a) Illustration of a trajectory evolving in a system with $N=2$ spins corresponding to one spin dimer.  The systems ``jumps'' from sheet $S_1$ to $S_2$, when reaching the threshold value $\alpha_1 = q^*_{12}$ and vice versa at $\beta_1 = q^*_{21}$. The Poincar\'{e} section in the $q$-$v$-space at $\phi=0$ is illustrated by $\textcolor{blau}{\boldsymbol{\Box}}$ and $\textcolor{rot}{\boldsymbol{\times}}$. (b) Poincar\'{e} section of an actual simulated trajectory with $J=2$, $p=2$, $b_1=2$ and $b_2=0$. In the bistable region one finds blue as well as red dots, which indicates an overlapping of two attractors living within $S_1$ and $S_2$.}
    \label{fig:N=2_schema}
\end{figure}

\textit{$N=2$}: In Fig.\,\ref{fig:N=2_schema}a) a schematic trajectory projected in the $q$-$v$-space is shown. In the case of $N=2$ there are two sheets $S_1$ and $S_2$ corresponding to the two ($ k = 0,1 $) different values of the magnetization $ M_0 = -1 $ and $M_1 = +1 $. In accordance to our discussion from section \ref{sec:twospins}, it can be seen, that there are two threshold values $ \alpha_1 = q^*_{12} $ and $ \beta_1 = q^*_{21} $ where the system ``jumps'' from one sheet to the other. In addition in Fig.\,\ref{fig:N=2_schema}a) we illustrated the Poincar\'{e} section $(q,v,\phi=0)$ in $S_1$ with $\textcolor{blau}{\boldsymbol{\Box}}$ and in $S_2$ with $\textcolor{rot}{\boldsymbol{\times}}$.
The Poincar\'{e} section for an actual simulated trajectory with $J=2$, $p=2$, $b_1=2$ and $b_2=0$ is also shown in Fig.\,\ref{fig:N=2_schema}b). We see, that in the bistable region $ -2 < q < 0 $ the Poincar\'{e} sections for $S_1$ and $S_2$ are overlapping. Hence, in case a chaotic attractor is projected into the $q$-$v$-subspace, the different ``parts'' of the attractor existing within different sheets could overlap, because of the multistability of $M$.  

\begin{figure}
    \centering
    \includegraphics[width=1.0\columnwidth]{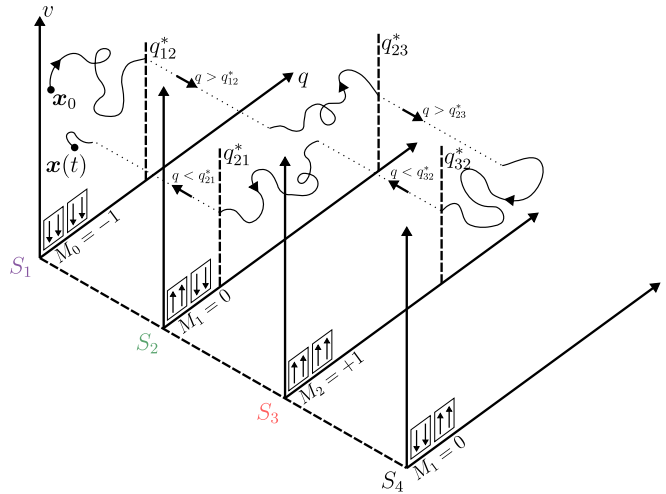}
    \caption{Schematic trajectory evolving in the phase space given by $\{ \, q,v,\Omega t\,\mbox{mod}\,2\pi \, \} \times \{ \, \pm 1 \, \}^{N/2}$, here for $N=4$. The discrete values of the spin dimer configurations are illustrated by different sheets $S_1,\dots,S_4$. The full description is given in the text.}
    \label{fig:N=4_schema}
\end{figure}

\textit{$N=4$}: Again in Fig.\,\ref{fig:N=4_schema} a schematic trajectory projected into the $q$-$v$-space is shown. In the case of two spin dimers, there are 4 different sheets $S_1,\dots,S_4$. Each sheet corresponds to one spin dimer configuration illustrated by the two boxes with arrows. The related values of the magnetization are also illustrated by $M_0 = -1$ ($S_1$), $M_1=0$ ($S_2,S_3$) and $M_2 = +1$ ($S_4$). Also for two spin dimers there are four threshold values, where a spin dimer flips. Here we will assume without loss of generality, that the upper threshold value $\alpha_1 = q^*_{12}$ of the left (``first'') spin dimer in Fig.\,\ref{fig:N=4_schema} is lower then the upper value of the right (``second'') dimer $\alpha_1 < \alpha_2 = q^*_{23} $. Imagine a arbitrary trajectory starting at $\vc{x}_0$ in sheet $S_1$. If the position $q$ is greater than the upper threshold value of the first dimer $\alpha_1 = q^*_{12}$, the first dimer flips from its down to the up state and the trajectory evolves within the sheet $S_2$. Again, if $q$ becomes greater then $ \alpha_2 = q^*_{23} $ the second dimer flips up and the system stays in $S_3$. Because we assume the same width of the hysteresis loop of each dimer we have $ \beta_1 = q^*_{21} < \beta_2 = q^*_{32} $, hence if $q$ becomes lower then the threshold $ \beta_2 = q^*_{32} $ the same second dimer changes its orientation again. The trajectory evolves in $S_2$ again. Finally if $ q < \beta_1 = q^*_{21} $ the first dimer flips back to the down state and the system reached $S_1$. 

Note, that because we assume $p$ to be constant, for an increasing input followed by a decreasing one, the spin dimers are switching from the down to the up and back to the down state by exactly the same reversed order. Thus, out of the $2^{N/2}$ sheets not more than $N/2+1$ can be reached. So each sheet corresponds to only one spin state and to one unique magnetization value $M_k$. This also means, that the spin states of the system can be determined by the actual input-output values $(q,M)$, without the knowledge of the history of the system. Furthermore it can be seen, that in the case of the thermodynamic limit (see Sec.\,\ref{sec:limit}) for a given point $(q,M)$ the slope of the hysteresis curve is unique and no intersections are possible. Therefore the system shows local memory \cite{Mayergoyz1991a}.

\begin{figure}
    \centering
    \includegraphics[width=0.95\columnwidth]{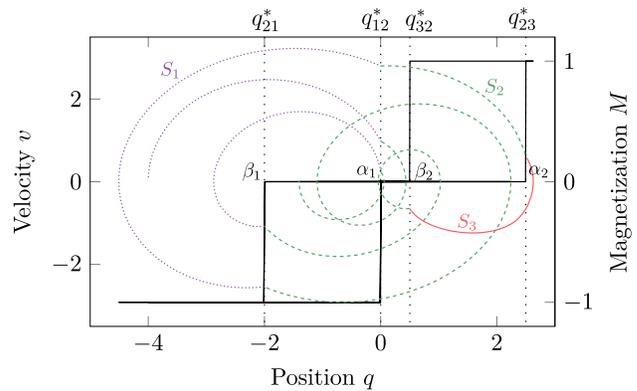}
    \caption{Simulation of an actual trajectory with $J=2$, $p=2$, $b_1=2$, $b_2=0$, $b_3=-0.5$ and $b_4=-2.5$. The different sheets of the phase space are labeled with $S_1$ (\textcolor{lila}{\linie}, dotted), $S_2$ (\textcolor{gruen}{\linie}, dashed) and $S_3$ (\textcolor{lachs}{\linie}, solid).}
    \label{fig:N=4_traj}
\end{figure}

In Fig.\,\ref{fig:N=4_traj} we also simulated an actual trajectory and coded the different sheets by different colors and styles: $S_1$ (\textcolor{lila}{\linie}, dotted), $S_2$ (\textcolor{gruen}{\linie}, dashed) and $S_3$ (\textcolor{lachs}{\linie}, solid). Here the following parameters were chosen: $J=2$, $p=2$, $b_1=2$, $b_2=0$, $b_3=-0.5$ and $b_4=-2.5$. For the threshold values we find in accordance with the simulation $ \alpha_1 = 0 $, $\beta_1 = -2 $, $ \alpha_2 = 2.5 $ and $ \beta_2 = 0.5 $.

\textit{Lyapunov exponents}: Now we want to describe the calculation of the Lyapunov exponents for a system like Eq.\,\eqref{eq:eom_4}. From Eq.\,\eqref{eq:eom_4} we can see, that our system behaves regular within the regions $S_n$ of the phase space. In this case the system is a damped harmonic oscillator with periodic driving force and an additional constant force $F_{\text{hys}}=C M_k$. Hence, chaotic behavior can only be introduced by intersections of the trajectories with boundaries between smooth regions. This means, that specific attention must be paid for the calculation of the Lyapunov exponents at these intersections. 

In general Lyapunov exponents are defined as the average rate of divergence or convergence between a reference trajectory $ \vc{x} $ and a perturbed trajectory $ \tilde{\vc{x}} = \vc{x} + \delta \vc{x} $. Within the smooth region the standard method for calculating Lyapunov exponents can be used \cite{1980bennetin}. In case the reference trajectory crosses a discontinuity boundary, the determination of the dynamic behavior of a infinitesimal perturbation $\delta \vc{x}$ becomes more complicated. This is because of the fact, that the reference trajectory and the perturbed trajectory does not reach the intersection point at the same time, but with time lag $\delta t$. To correct this difference in the switching behavior of $\vc{x}$ and $\tilde{\vc{x}}$, the concept of \ac{DM} can be applied \cite{nordmark1991,mueller1995,dankowicz2000,bernardo20011}. If we assume, that $ \vc{x} $ crosses the boundary before $ \tilde{\vc{x}} $ with a small time lag $\delta t$, then we can introduce the discontinuity map $ \vc{Q} $, which maps $ \tilde{\vc{x}} $ from before to after its crossing, at the moment $ \vc{x} $ crosses the boundary. Thereby $ \vc{Q} $ is taking the effect of the discontinuity into account. Thus, the \ac{DM} incorporates the effect of a discontinuity crossing even if the state is only in the neighborhood of a boundary and the crossing appears in the future.
For our system the \ac{DM} is given by
\begin{equation}
    \tilde{\vc{x}} \to \vc{Q}(\tilde{\vc{x}}) = 
    \begin{pmatrix}
    \tilde{q} \\
    C (\Delta M)\delta t + \tilde{v} \\
    \tilde{\phi}
    \end{pmatrix},
\end{equation}
where $\delta t=\frac{\tilde{q}-q^*}{v^*}$. Here $ q^* $ is the intersection point with the boundary, $ v^* $ is the velocity at intersection and $ \Delta M $ is the change of the magnetization across the boundary. 

Then, for infinitely small perturbations $ \delta \vc{x} $ the Jacobian $ \mt{X} = \pdv{\tilde{\vc{x}}}\vc{Q}(\tilde{\vc{x}}) $ of the map $ \vc{Q}(\tilde{\vc{x}}) $ can be used to calculate the perturbation $ \delta \vc{x}^+(t^*)$ after intersecting the discontinuity boundary $ \delta \vc{x}^+(t^*) = \mt{X} \delta \vc{x}^-(t^*) $ from the perturbation directly before the boundary crossing $\vc{x}^-(t^*)$. The matrix $ \mt{X} $ is often called saltation matrix. For our system we find
\begin{equation}
    \mt{X} = \begin{pmatrix}
                     1 & 0 & 0 \\
                 \frac{1}{v^*} C \Delta M & 1 & 0 \\
                0 & 0 & 1
                \end{pmatrix} . \label{eq:salt_matrix}
\end{equation}
With the saltation matrix $\mt{X}$, the largest Lyapunov exponent for a reference trajectory with only one crossing at time $t^*$ can be written as 
\begin{equation}
\label{eq:lyap-ex}
    \lambda_\text{max} = \lim_{t \to \infty} \frac{1}{t} \ln \frac{| \mt{Y}(t,t^*) \mt{X} \mt{Y}(t^*,0) \delta \vc{x}_0|}{|\delta \vc{x}_0|},
\end{equation}
where $ \mt{Y}(t,t') $ is the fundamental solution of the variational equation $ \delta \vc{\dot x}(t) = \mt{J}_{\vc{x}} \, \delta \vc{x}(t) $ of Eq.\,\eqref{eq:eom_4} from time $t'$ to $t$ and $\mt{J}_{\vc{x}}$ denotes the Jacobian matrix. In general, at each crossing a multiplication with the corresponding saltation matrix is necessary and we use a QR-decomposition to calculate the Lyapunov exponents for large $t$. A more detailed description can be found in \cite{zech2020}.

To use the concept of \ac{DM} for the complex hybrid phase space described in Sec.\,\ref{sec:pws} it is useful to consider the system in a different way. Therefore we will treat the system as it is evolving in the continuous phase space $\{ \, q,v,\phi \,\mbox{mod}\,2\pi \, \}$, so that the Lyapunov exponents are well defined. This phase space consist of the regions $S_n$, which are separated by the boundaries $q^*_{nm}$. But in contrast to the case of independent spins, which we investigated in \cite{zech2020}, here to positions of the boundaries change over time. Thus some of the boundaries are ``active'' or ``inactive'' over time. In the next section we will describe how to use this point of view to simulate trajectories of this piecewise-smooth system.

\label{sec:numerics}
\textit{Numerics}: Since the disorder values $b_i$ are quenched and the spins are pairwise coupled to each other, we are able to calculate a priori the solution $x(t)$ at all possible discontinuities, where a spin flips can occur. For the generation of a trajectory of the piecewise-smooth system we start with an initial condition at $ t = 0 $ with the corresponding initial magnetization $M_k$ and use the analytical solution of the linear system \eqref{eq:eom_4}, which is described in \cite{zech2020}. By discretizing this solution with time steps $ \Delta t $ we propagate the trajectory until the first spin flip occurs, where the magnetization jumps from the initial value $M_k$ to some other value. At all potential spin flips we determine the metastable state of the spin dimer system to check whether this boundary is an ``active'' boundary, i.\,e., whether a spin flip occurs or not. This depends on the history of the system. If it is indeed an intersection point with a jump of the magnetization we use a root finder to exactly determine the time of the discontinuity. After the boundary we can simply use the analytic solution with the new magnetization value to propagate the trajectory within the next smooth region of the phase space. 

In case of a large number of spins the sizes of the smooth regions in the phase space are very small and the algorithm based on the analytical solution becomes very slow. For increasing the speed of the algorithm in this case we use a linearization of the solution to calculate directly the time to the next boundary, therefore in this case no root finder is necessary. 

By using the fundamental solution of the variational equation of Eq.\,\eqref{eq:eom_4} to evolve the perturbation between two boundaries and the saltation matrix from Eq.\,\eqref{eq:salt_matrix} at a boundary, we can calculate the corresponding Lyapunov spectrum by using a standard QR decomposition algorithm \cite{geist1990,dieci1997}.

\subsection{Preisach model}
\label{sec:preisach}

The harmonic oscillator coupled to pairwise interacting Ising spins results in a system with $ N/2 $ superposed elementary hysteresis loops, which resembles the definition of the \ac{PM}. Thus we want to give a quick overview of the basic ideas of the \ac{PM}. The building blocks of the \ac{PM} \cite{Preisach1935,Mayergoyz2003} are the elementary hysteresis loops, which are also called Preisach units, hysterons or relays. An example of such an elementary hysteresis loop is given by the blue solid curve (\textcolor{blau}{\linie}) in Fig.\,\ref{fig:elem_hys}. For a given input $ q(t) $ the output $ y_{\alpha \beta}[q](t) $ of a Preisach unit at time $t$ with threshold values $\alpha$ and $\beta$ ($\alpha>\beta$) is given by
\begin{equation}
    y_{\alpha \beta}[q](t) = \begin{cases}
                +1  & \text{for \;} q(t_0) \ge \alpha \text{\;and\;} q(t_1) \ge \beta, \, \forall \, t_1 \in [t_0,t], \\
                 -1 & \text{for \;} q(t_0) < \beta \text{\;and\;} q(t_1) < \alpha, \, \forall \, t_1 \in [t_0,t],
                \end{cases} \label{eq:preisachunit}
\end{equation}
where $t_0$ specifies the last time at which the input was outside the bistable interval $[\beta,\alpha]$. Since, the time $t_0$ depends on the behavior of the position function $q(t')$ with $t'<t$, the output $y_{\alpha \beta}[q](t)$ of the Preisach units is, in general, a functional. The output of the whole \ac{PM} is a superposition of the output of infinitely many Preisach units with different threshold values weighted by the so-called Preisach density $ \mu(\alpha,\beta) $. For our ferromagnetic interpretation of the complex hysteresis, the output of the \ac{PM} is the magnetisation $ M[q](t) $, which is defined as
\begin{equation}
  M[q](t) = \iint\limits_{\alpha \ge \beta} \mu(\alpha,\beta) y_{\alpha \beta}[q](t) \, \mathrm{d}\alpha \, \mathrm{d}\beta . \label{eq:preisachop}
\end{equation}

The parameters $ \alpha $ and $ \beta $ span a surface, the so-called Preisach plane, which can be used to illustrate the Preisach density or the internal state of the \ac{PM}. However, since $\alpha>\beta$ only a half-plane in the two-dimensional Preisach plane is relevant. A very common transformation of the Preisach plane is a rotation by $+\pi/4$ and a scaling by a factor $1/\sqrt{2}$, where the new variables $s$ and $r$ are specified by
\begin{equation}
    s = \frac{\alpha + \beta}{2}, \quad r = \frac{\alpha-\beta}{2} .
    \label{eq:transform}
\end{equation}
The variable $s$ describes the center and $r$ is the half of the width of an elementary hysteresis loop ($r>0$).

In the new coordinates the relationship between the \ac{PM} and the ensemble of spin dimers becomes clear, by identifying corresponding elementary hysteresis loops from each system: the values $\alpha_k$ and $\beta_k$ from Eq.\,\eqref{eq:jump_val} in Sec.\,\ref{sec:manyspindimers} are identical to the threshold values captured by the Preisach density of Eq.\,\eqref{eq:preisachop}. For a finite ensemble of spin dimers the Preisach density is given by the discrete density $\mu(\alpha, \beta) = \sum_k^{N/2} \del(\alpha-\alpha_k) \, \del(\beta-\beta_k)$, whereas a continuous Preisach density $\mu(\alpha, \beta)$ corresponds to the thermodynamic limit of the spin system. For instance, if we consider the example from Sec.\,\ref{sec:manyspindimers}, where we assumed that the center $ s $ of the elementary hysteresis loops is Gaussian distributed with $ s \sim \mathcal{N}(0,R^2) $ and the width of the each loop is fixed equal to $p$, the corresponding transformed Preisach density $\tilde{\mu}(r,s)$ given as
\begin{equation}
\label{eq:pdens_sr}
\tilde{\mu}(r,s) = \mu(s) \, \del(r-\frac{p}{2}) ,
\end{equation}
with 
\begin{equation}
 \mu(s) = \frac{1}{\sqrt{2 \pi R^2}} \E^{-\frac{s^2}{2 R^2}} . \label{eq:gauss}   
\end{equation}
This relationship between the system with pairwise interacting spins and the Preisach model is especially helpful for calculating the magnetization if the number of spin dimers goes to infinity.

\subsection{Play operator and thermodynamic limit}
\label{sec:limit}
From Eq.\,\eqref{eq:pdens_sr} one can see, that the width of the elementary hysteresis loops are delta distributed and the center is Gaussian distributed in the Preisach plane. It has been shown, that when using a uniformly distributed center of the loops, the output of the \ac{PM} is given by the so called ``play'' or ``backlash'' operator \cite{Brokate1996,dimian2013}. The output $ w(t) $ of this operator at time $t$ is given by:
\begin{equation}
    w(t) = \operatorname{max} \biggl\{ q(t)-\frac{p}{2} ,\operatorname{min} \biggl\{ q(t) + \frac{p}{2}, w(t_i) \biggr\} \biggr\} ,
    \label{eq:real_play}
\end{equation}
where $t_i < t$ is the time when the last extremum of the input $q$ was attained.

\begin{figure}
    \centering
    \includegraphics[width=0.95\columnwidth]{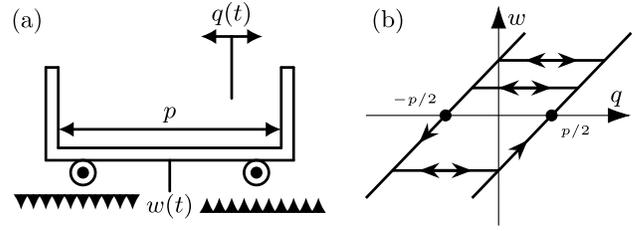}
    \caption{(a): Illustration of mechanical play by using the analogue of a wagon. The input $q(t)$ is given by e.\,g. a finger, which moves the wagon. If the finger touches the right or the left side of the wagon, respectively, the output is given by $ w(t) = q(t) - p/2 $ and $ w(t) = q(t) + p/2 $, whereas the output stays constant if the finger moves in between both sides of the wagon. The black arrows on the left and right side of $w$ illustrate the separation of the spin dimers (ordered according to the center values of their associated hysteresis loops) in the down or up state, respectively. (b): Representation of the movement of the wagon in the input-output-plane.}
    \label{fig:play_mech}
\end{figure}

Since $t_i$ depends on the history of the input, $w(t)$ is a functional of the input function $q(t')$ with $t'<t$. The expression in Eq.\,\eqref{eq:real_play} can be visualized by using a mechanical analogue consisting of a wagon controlled by a finger movement (see Fig.\,\ref{fig:play_mech}(a)). Here the input $q(t)$ is the position of a finger that moves a wagon. The center $ w(t) $ of the wagon only changes, if the finger position during a movement is identical with the right ($w(t) = q(t)-\frac{p}{2}$) or the left position ($w(t) = q(t)+\frac{p}{2}$) of the wagon, respectively. In case the finger stands in between both walls, $ w(t) $ does not change and is identical to its last position in a movement. The corresponding behavior in the input-output-plane is illustrated in Fig.\,\ref{fig:play_mech}(b).

\begin{figure}
    \centering
    \includegraphics[width=0.95\columnwidth]{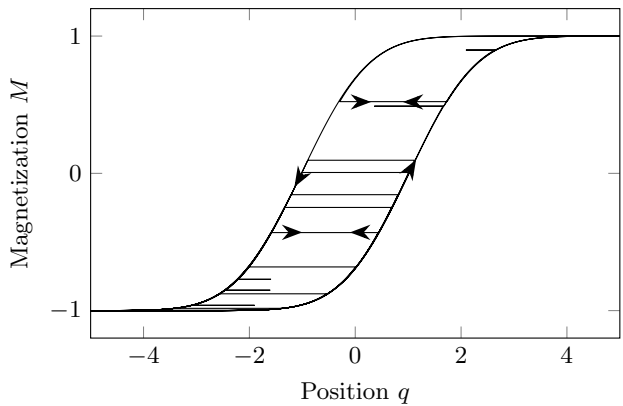}
    \caption{Illustration of the output of the spin dimer system in its thermodynamic limit, resulting in an operator in form of a generalized mechanical play ($p=2$, $R=1$) given by Eq.\,\eqref{eq:play}. The input-output behavior is characterized by a major loop formed by the functions $f_-$ and $f_+$ from Eqs. \eqref{eq:play_1}, \eqref{eq:play_2} and degenerate inner loops.}
    \label{fig:play}
\end{figure}

Because each spin dimer in our system contributes one elementary hysteresis loop, we can make a connection between the mechanical play and the the system of spin dimers: The input $q(t)$ corresponds to the position of the finger in the mechanical play and to the value of the external field in the spin system. Now one imagines the dimers placed on the $q$-axis according to their center values $s_k$ (see \ref{sec:manyspindimers}), i.\,e. in creasing order from left to right (the arrows in Fig.\,\ref{fig:play_mech}(a). Then a movement of the wagon (with center position at $w(t)$) to the right (left) corresponds to an increasing (decreasing) field, which flips dimer states centered at position $w(t)=q(t)-p/2$ from $-1$ to $+1$ ($w(t)=q(t)+p/2$ from $+1$ to $-1$). If the finger moves inside the wagon without touching the walls the position $w$ of the wagon stays at its last position of a movement. This corresponds to a variation of the external field within the elementary hysteresis loop which switched at last. Thus the position $w$ of the wagon separates the dimers in the down-state from the dimers in the up-state. For this to hold we assume, that initially all dimers were in the down-state and the external field assumed its minimal value. In Fig.\,\ref{fig:play_mech} we illustrated this by the arrows pointing down on the left side of $w$ and pointing up on the right side, respectively. Hence the magnetization in the thermodynamical limit can be calculated by summing over all dimer states or, in the thermodynamic limit by an integration over the density of the centers of the elementary hysteresis loops, given by $\mu(s)$ from Eq.\,\eqref{eq:gauss}:
\begin{equation}
M[q](t) = 2  \int\limits_{-\infty}^{w(t)} \mu(s) \, \mathrm{d}s - 1 = \operatorname{erf}\left( \frac{w(t)}{\sqrt{2}R} \right) , \label{eq:prob_mag}
\end{equation}
where $ w(t) $ is given by Eq.\,\eqref{eq:real_play}. Using the monotonicity of the error function this can also be written in a more compact way:
\begin{equation}
    M[q](t) = \operatorname{max} \biggl\{ f_{+}(q(t)),\operatorname{min} \biggl\{ f_{-}(q(t)),M[q](t_i) \biggr\} \biggr\} \label{eq:play} ,
\end{equation}
with
\begin{align}
    f_{-}(q(t)) &= \operatorname{erf}\left( \frac{q(t) + \frac{p}{2}}{\sqrt{2}R} \right) , \label{eq:play_1} \\
    f_{+}(q(t)) &= \operatorname{erf}\left( \frac{q(t) - \frac{p}{2}}{\sqrt{2}R} \right) . \label{eq:play_2} 
\end{align}
This form of the hysteretic play operator can be found in the literature under the term ``generalized play'' \cite{Visintin1994,krasnosel1989,HASSANI2014}.
In Fig.\,\ref{fig:play} an illustrative case with $ p = 2 $ and $ R = 1 $ is shown. Note, that Eq.\,\eqref{eq:play} can also be derived by integrating Eq.\,\eqref{eq:preisachop} with a Preisach density $\mu(\alpha,\beta)$ obtained from Eqs. \eqref{eq:pdens_sr} and \eqref{eq:gauss} using the transformation from Eq.\,\eqref{eq:transform}.

\begin{figure}
    \centering
    \includegraphics[width=1.0\columnwidth]{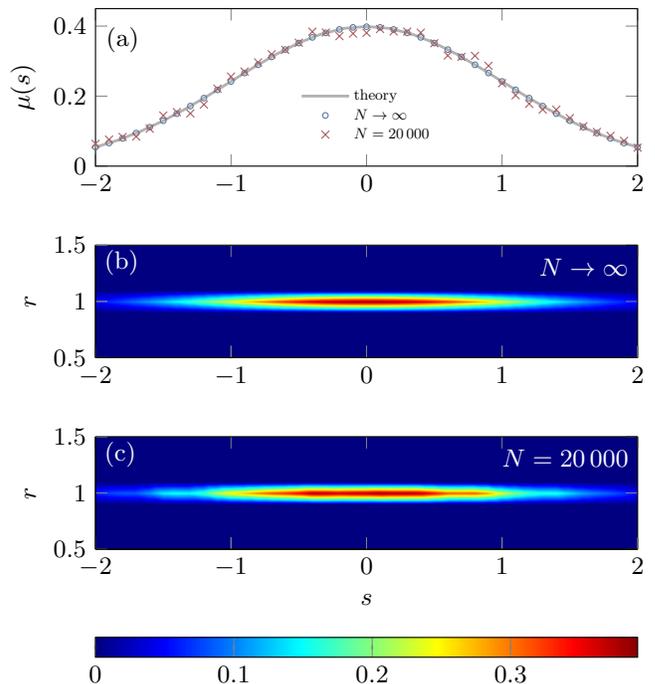}
    \caption{(a): Comparison of the numerically calculated and theoretical density $\mu(s)$ from Eq.\,\eqref{eq:gauss}. (b), (c): The numerically determined Preisach density plotted in the $r$-$s$-plane. $ N \to \infty $ denotes the system in its thermodynamic limit and $ N=20\,000 $ denotes the piecewise-smooth system. The randomness and the width of each elementary hysteresis loop are set to $R=1$ and $p=2$.}
    \label{fig:preis_dens}
\end{figure}

As a verification of our calculations we determined the Preisach density $ \mu(r,s) $ of the piecewise-smooth system for $ N=20\,000 $ spin dimers with randomness $R=1$, as well as for the system in its thermodynamic limit, given by Eq.\,\eqref{eq:play}. A detailed description how to determine $ \mu(r,s) $ in a experimental way by scanning the whole Preisach plane gradually can be found in \cite{Mayergoyz2003,urbanaviciute2018}. The results are shown in Fig.\,\ref{fig:preis_dens}. We find, that for both systems the density coincides within numerical accuracy with the result from Eqs. \eqref{eq:pdens_sr} and \eqref{eq:gauss}. 

\textit{Lyapunov exponents}: Like mentioned before, in general, the largest Lyapunov exponent $ \lambda_\text{max} $ specifies the average exponential behavior of an infinitesimal perturbation of a reference trajectory. The time evolution of the infinitesimal perturbation can be described by the linearization of Eq.\,\eqref{eq:eom_2}. Note, that the r.h.s of Eq.\,\eqref{eq:eom_2} depends on the functional $M[q](t)$ given by Eq.\,\eqref{eq:prob_mag}, which depends on the output $w(t)$ (the wagon position) of the play operator \eqref{eq:real_play}. Therefore, it is necessary to consider the extended state space $\vc{y}=(q,v,\phi,w)^T$ (see details in \ref{sec:app}), which also stores the memory of the system, in form of the position of the wagon. The time evolution of $\vc{y}$ can be given by 
\begin{equation}
\dot{\vc{y}}(t) =
     \begin{pmatrix}
                v(t) \\
                 - 2\zeta v(t) - q(t) + \cos \phi(t) + g(t,t_i') \\
                \Omega \\
                \chi \, v(t)
                \end{pmatrix}  \label{eq:eom_smooth_4}
\end{equation}
with
\begin{equation}
   g(t,t_i') = C \bigl( \chi M(w(t)) + (1-\chi) M(w(t_i') \bigr) .
\end{equation}
Here the function $\chi(\vc{y}(t))$ indicates whether the state $\vc{y}(t)$ corresponds to a situation ``inside'' or ``outside'' of the play
\begin{equation}
    \chi(\vc{y}(t)) = \begin{cases}
            0,  & \text{for } |w(t) - q(t)| < p/2, \text{ inside} , \\
            1,  & \text{for } |w(t) - q(t)| = p/2, \text{ outside} 
            \end{cases}
            \label{eq:indicator}
\end{equation}
and $ t_i' $ is the time when the last extremum of the input $q$ was attained on condition that at this time the system changes from ``outside'' to ``inside'' of the play. Hence $ t_i' $ is the last time, when the finger left one of the walls of the wagon from Fig.\,\ref{fig:play_mech}(a). Therefore extrema, which only occur ``inside'' of the play, are not denoted by $ t_i' $. 

For infinitesimal perturbation $\del \vc{y}(t)$ of a reference solution $\vc{y}^R(t)$, we can distinguish basically between two situations. For $\chi(\vc{y}^R(t))=0$ ("inside" the play) the Jacobian is given by
\begin{equation}
  \mt{W}^0(t,t_i') = \begin{pmatrix}
     0 & 1 & 0 & 0 \\
     -1 & -2 \zeta & - \sin \phi^R(t) & C \left. \frac{\pd M(w)}{\pd w} \right|_{w^R(t_i')} \\
     0 & 0 & 0 & 0 \\
     0 & 0 & 0 & 0
    \end{pmatrix} , 
    \label{eq:limit_W_0}
\end{equation}
whereas for $\chi(\vc{y}^R(t))=1$ (outside), we have
\begin{equation}
  \mt{W}^1(t) = \begin{pmatrix}
     0 & 1 & 0 & 0 \\
     -1 & -2 \zeta & - \sin \phi^R(t) & C \left. \frac{\pd M(w)}{\pd w} \right|_{w^R(t)} \\
     0 & 0 & 0 & 0 \\
     0 & 1 & 0 & 0
    \end{pmatrix}  .
    \label{eq:limit_W_1}
\end{equation}
Hence, the variational equation of $ \vc{y} $ can be written as $ \del \dot{\vc{y}}(t) = \mt{W}^\chi \del \vc{y}(t) $.

First we note, that for both matrices the entries $W_{ij}$ with $i,j=1,\dots,3$ are similar and correspond to the Jacobian of a damped harmonic oscillator with periodic forcing and without play operator. In the following we want to explain the differences between $ \mt{W}^0 $ and $ \mt{W}^1 $, starting with $ \mt{W}^1 $ (outside): Here the finger touches one of the walls of the wagon (see Fig.\,\ref{fig:play_mech}(a)) and therefore the actual position of the wagon is given by $ w(t) = q(t) \pm p/2 $, which means there is no memory in the system. Thus the time evolution of the corresponding perturbations $ \del q $ and $ \del w $ should behave in a similar way ($\del \dot{q} = \del v$, $\del \dot{w} = \del v$). This is satisfied by the fact, that the first and the last line of $ \mt{W}^1(t) $ in Eq.\,\eqref{eq:limit_W_1} are identical. Note, that both matrices also differ in the entry $ W_{24} $. Here $ W^1_{24} $ gives the linearization of Eq.\,\eqref{eq:play} evaluated at $ w^R(t) $. Because in this case we are outside of the play, from Eq.\,\eqref{eq:prob_mag} we simply get
\begin{equation}
   C \left. \frac{\pd M(w)}{\pd w} \right|_{w^R(t)} = C \sqrt{\frac{2}{\pi R^2}} \E^{-\frac{w^R(t)^2}{2 R^2}} .
\end{equation}
This means the system behaves like a harmonic oscillator with an additional nonlinear force without memory, which is indeed similar to the system we investigated in \cite{zech2020}. 

The situation changes rapidly, if the finger moves inside the wagon and $ w $ stays constant, storing the last extremum of the input $q$ at time $ t_i' $. Hence $ \del w $ does not evolve in the same way as $ \del q $, in fact $ \del w $ stays constant ($\del \dot{w}(t) = 0$), because actual perturbations of $ q $, $ v $ and $ \phi $ do not change the stored perturbation $ \del w $ at the time $ t_i' $. Furthermore, now $ W^0_{24} $ gives the linearization of Eq.\,\eqref{eq:play} evaluated at $ w^R(t_i') $. This means inside of the play $ \del v(t) $ is determined by $ \del w(t_i') $, when the system enters the play, which of course is an after-effect of the memory in the system.

Special attention has to be paid to the scenarios, when the system changes from $ \chi^R = 0 $ to $ \chi^R = 1 $ and vice versa. This change is described by crossing the boundary $|w^R(t) - q^R(t)| = p/2$ given by Eq.\,\eqref{eq:indicator} at the intersection point $ \vc{y}^* $ at time $t^*$. Note, that in case the system changes from $ \chi^R = 1 $ to $ \chi^R = 0 $, we have $t^* = t_i'$. By using the theory of piecewise-smooth system from section \ref{sec:pws} we can again determine a saltation matrix, which gives us $ \del \vc{y} $ immediately after crossing the boundary. Lets first deal with the case when the system enters the play ($\chi^R=1$ to $\chi^R=0$): Because here $q(t)$ is an extremum, $v(t)$ in Eq.\,\eqref{eq:eom_smooth_4} becomes zero and therefore the corresponding saltation matrix $ \mt{V}^\text{in} $, calculated in the usual way (see \cite{bernardo2008}) is simply $ \mt{V}^\text{in} = \mt{I} $, the four-dimensional identity matrix. This is clear, because in case the system enters the play, $ w(t) $ and its time derivative are smooth. This means $ \del w^1 $ immediately before and $ \del w^0 $ immediately after the transition from the outside to the inside of the play are identical $ \del \vc{y}^0(t^*) = \mt{V}^\text{in} \del \vc{y}^1(t^*) = \del \vc{y}^1(t^*) $. In contrast, when the system leaves the play $ \dot{w}(t) $ has a jump. This can be explained by using Fig.\,\ref{fig:play_mech}(a) again. Here at first the wagon holds its position with $ \dot{w} = 0 $. After the finger hits one of the walls, the new velocity $ v(t) $ is applied, hence $ \dot{w}(t) $ has a jump, because in this case $ q(t) $ is not an extremum ($ v(t) \ne 0 $) \footnote{In fact, if $ q(t) $ is an extremum (the finger touches one of the walls with velocity zero) this corresponds to a grazing intersection, which needs indeed to be investigate in more detail.}. Then the perturbation of $ \vc{y} $ after the transition is given by $ \del \vc{y}^1(t^*) = \mt{V}^\text{out} \del \vc{y}^0(t^*) $, with:
\begin{equation}
  \mt{V}^\text{out} = \begin{pmatrix}
     1 & 0 & 0 & 0 \\
     0 & 1 & 0 & 0 \\
     0 & 0 & 1 & 0 \\
     1 & 0 & 0 & 0
    \end{pmatrix}  .
    \label{eq:salt_matrix_limit}
\end{equation}
This is quite remarkable, because the last line of $ \mt{V}^\text{out} $ shows, that when leaving the play region the perturbation $ \del w $ is been wiped out by $ \del w^1 = \del q^0 $. This leads to repeatedly dimensional collapses from the four-dimensional system to only three dimensions. Hence one corresponding Lyapunov exponent becomes minus infinity. This behavior typically can found in systems with time- or state-dependent delays \cite{radons2009,otto2010,wang2011}.

\textit{Numerics}: Finally the Lyapunov spectrum can be calculated in a similar way as described in Sec.\,\ref{sec:numerics}. Here we discretize the system with time steps $ \Delta t $. Inside of the play the discretization of Eq.\,\eqref{eq:limit_W_0} can be used to determined the evolution of the perturbation, whereas outside of the play we can use Eq.\,\eqref{eq:limit_W_1}. In case the system undergoes a transition from the inside of the play to the outside of the play we use the saltation matrix from Eq.\,\eqref{eq:salt_matrix_limit} to correct the perturbations. By applying this concept we calculate the whole Lyapunov spectrum by using again a QR decomposition algorithm. 

\textit{Fractal Dimensions}: By ordering all $ k $ Lyapunov exponents from largest to smallest $\lambda_1 = \lambda_\text{max} \ge \lambda_2 \ge \dots \ge \lambda_4 $ we are able to calculate the so-called 
Kaplan-Yorke dimension of a chaotic attractor, also known as Lyapunov dimension, which is defined as \cite{grassberger1983,kaplan1979}:
\begin{equation}
    D_\text{KY} = j + \frac{\sum_{i=1}^j \lambda_i}{|\lambda_{j+1}|} , \label{eq:kaplan_yorke}
\end{equation}
where $ j \in \mathbb{N}: \sum_{i=1}^j \lambda_i \ge 0 \land \sum_{i=1}^{j+1} \lambda_i < 0 $. 

This means that $j$-dimensional volumina are still expanding (with rate $\sum_{i=1}^{j}\lambda _{i}$), whereas $(j+1)$-dimensional volumina are contracted with the rate $\sum_{i=1}^{j+1}\lambda_{i}$. The second part of Eq.\,\eqref{eq:kaplan_yorke} follows from a linear interpolation between these two cases such that expansion and contraction balance yielding a zero rate for volumina with (fractal) dimension $D_\text{KY}$.

Below we will also determine the box counting dimension $ D_0 = D_\text{BC} $. The box counting dimension of a chaotic attractor can be determined by partitioning the phase space of the attractor into a grid of boxes with size $ \epsilon $ and a subsequent counting of the number $ N $ of boxes, which contain points. Thus the scaling with the grid size defines the box counting dimension:
\begin{equation}
    D_0 = D_{BC} = \lim_{\epsilon \to 0} \frac{\log N(\epsilon)}{\log 1/\epsilon} \label{eq:box_counting}.
\end{equation}
The Kaplan–Yorke conjecture states, that the Kaplan-Yorke dimension equals the information dimension $ D_1 $ (which is lower than the box counting dimensions $ D_\text{BC} \ge D_\text{KY} $) for ``typical'' systems \cite{kaplan1979}. We are not calculating the information dimension $ D_1 $, but it is worth to mention, that the main difference between the box counting dimension $ D_0 $ and $ D_1 $ is given by the fact, that in contrast to $ D_0 $ by calculating $ D_1 $ the amount of points found in one box of the scaled grid is taken into account. For a more detailed view on the topic of fractal dimensions see \cite{farmer1983}.    
    
\section{Results} \label{sec:results}
   Here we present our numerical results. If not stated otherwise, for the numerical simulations we used the initial conditions and model parameters from Table\,\ref{tab:parameters} for the piecewise-smooth system as well as for the system in its thermodynamic limit.

\begin{table}[!ht]
    \begin{tabular}{|c|c|c|} \hline 
        Parameter & Symbol & Value \\ \hline \hline
        \multicolumn{3}{|c|}{General} \\ \hline
        damping ratio & $\zeta$ & 0.05 \\
        excitation frequency & $\Omega$ & 1.0 \\
        initial position & $q_0$ & -1.0 \\
        initial velocity & $v_0$ & 0.1 \\
        elementary hysteresis loop width & $p$ & 2.0 \\ 
        randomness & $R$ & 1.0 \\ \hline
        \multicolumn{3}{|c|}{Piecewise-smooth} \\ \hline
        initial spin orientation & $\sigma_i(t=0)$ & -1 \\ 
        spin coupling strength & $J$ & 2.0 \\ \hline
        \multicolumn{3}{|c|}{Thermodynamic limit} \\ \hline
        initial output of the play operator & $w_0=q_0-p/2$ & -2.0 \\ \hline
    \end{tabular}
    \caption{Model parameters and initial conditions of the numerical simulations for the piecewise-smooth system as well as for the system in its thermodynamic limit.}
    \label{tab:parameters}
\end{table}

In the following we want to distinguish between the fractal dimension of the piecewise-smooth system and the system in its thermodynamic limit by denoting the fractal dimensions either with $D^N$ or $D^\infty$. Also the corresponding Lyapunov exponents will be denoted in the same way.

For the piecewise-smooth system we find the three Lyapunov exponents $ \lambda_1^N = 0 $, $ \lambda_2^N $ and $ \lambda_3^N $, where the sum of all exponents equals the volume contraction of the corresponding phase space $ \sum_i \lambda_i = \vc{\nabla} \cdot \vc{F} = -2\zeta $. Here $ \vc{F} $ is the right-hand side of Eq.\,\eqref{eq:eom_4}. Thus, one can show, that for regular behavior we have $\lambda_1^N = 0 $,  $\lambda_2^N = -\zeta $ and  $\lambda_3^N = -\zeta $, which is in full accordance with our numerical findings (see e.\,g. Fig.\,\ref{fig:bif_0}). Note, that one of the three Lyapunov exponent associated to perturbations of the phase $ \phi $ of the external forcing equals zero. However, for calculating $ D_\text{KY}^N $ these perturbations are neglected in the following discussions and thus the system is considered within the Poincar\'{e}-section $\phi = 0$. Furthermore instead of calculating $ D_\text{BC}^N $ for each sheet $S_n$ we can determine $ D_\text{BC}^N $ from the superposition of all $N/2+1$ sheets in the $q$-$v$-space, as long as the number of sheets is finite (see Fig.\,\ref{fig:N=2_schema}b)).

In case of the system in its thermodynamic limit we have four Lyapunov exponents  $ \lambda_1^\infty = 0 $, $ \lambda_2^\infty $, $ \lambda_3^\infty $ and $ \lambda_4^\infty = -\infty $. In the same way as for the piecewise-smooth system we want to neglect the direction of the perturbations associated to $ \phi $, when calculation $ D_\text{KY}^\infty $. As for the piecewise-smooth system, here the box counting dimension $ D_\text{BC}^\infty $ is also determined within the $q$-$v$-space. It is worth to mention, that this projection of the fractal attractor into the two-dimensional space equals the fractal dimensions of the three-dimensional space. This is because of the fact, that the projected fractal dimension are smaller than the dimension of the projective space \cite{falconer2013}, as we will see later.

\subsection{Bifurcations and attractors}

\begin{figure}
    \centering
    \def\svgwidth{\columnwidth}
    \includegraphics[width=1.0\columnwidth]{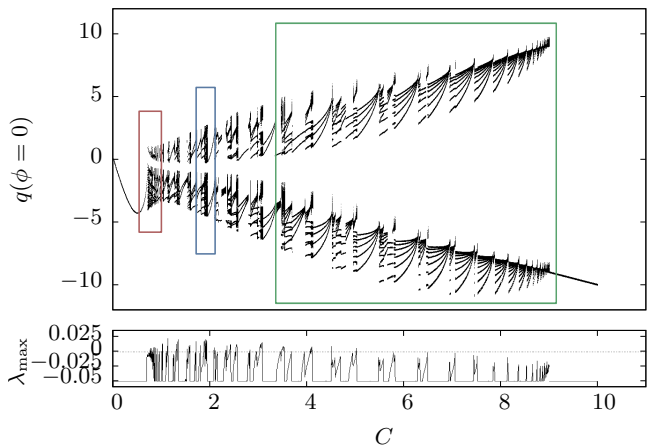}                         
    \caption{For two coupled spins $ N = 2 $ the system shows the bifurcation scenarios expected from piecewise-smooth square-root maps, illustrated by the different colored boxes: immediate jump to robust chaos with a positive largest Lyapunov exponent $\lambda_\text{max}$ (\textcolor{rot}{\linie}, left), period-adding with chaotic windows (\textcolor{blau}{\linie}, middle) and overlapping period-adding cascade (\textcolor{gruen}{\linie}, right). The local disorder fields are $ b_1 = 1 $ and $ b_2 = -1 $.}
    \label{fig:bif_0}
\end{figure}

\textit{Two spins}: At first, we study the basic system with a single dimer ($ N = 2 $). In this case we numerically determine the projection $q(\phi = 0)$ of the Poincar\'{e} section and the corresponding largest Lyapunov exponent $\lambda_\text{max} = \lambda_1$ in dependency on the coupling strength of the magnetization $ C $. The local field values are $ b_1 = 1 $, $ b_2 = -1 $. The corresponding bifurcation diagram is shown in Fig.\,\ref{fig:bif_0}. Starting with a large $ C $ and decreasing $ C $ step by step one finds, that at $ C = 9 $ the system undergoes a period-adding cascade (\textcolor{gruen}{\linie}, right). This is in accordance with the theoretic value of $ C $, which can be calculated by using the results from \cite{zech2020}. In the bifurcation diagram we can also find period-adding scenarios with chaotic regimes in between (\textcolor{blau}{\linie}, middle) and an immediate jump to chaos (\textcolor{rot}{\linie}, left). These are typical scenarios for piecewise-smooth systems with grazing behavior and piecewise-smooth square-root maps \cite{budd1994,chin1994,foale1994}. In \cite{zech2020} we derived the Poincar\'{e}-section discontinuity mapping and the zero-time discontinuity mapping for the system with a single spin, which indeed showed these kind of maps. Therefore it can be seen, that the system with only one spin and with two coupled spins behave in a similar way. It is worth to emphasize, that when changing the initial condition and/or the disorder realization the qualitative behavior of the bifurcations stays the same, but the position of the periodic windows and the chaotic regimes may change.  

\begin{figure}
    \centering
    \includegraphics[width=\columnwidth]{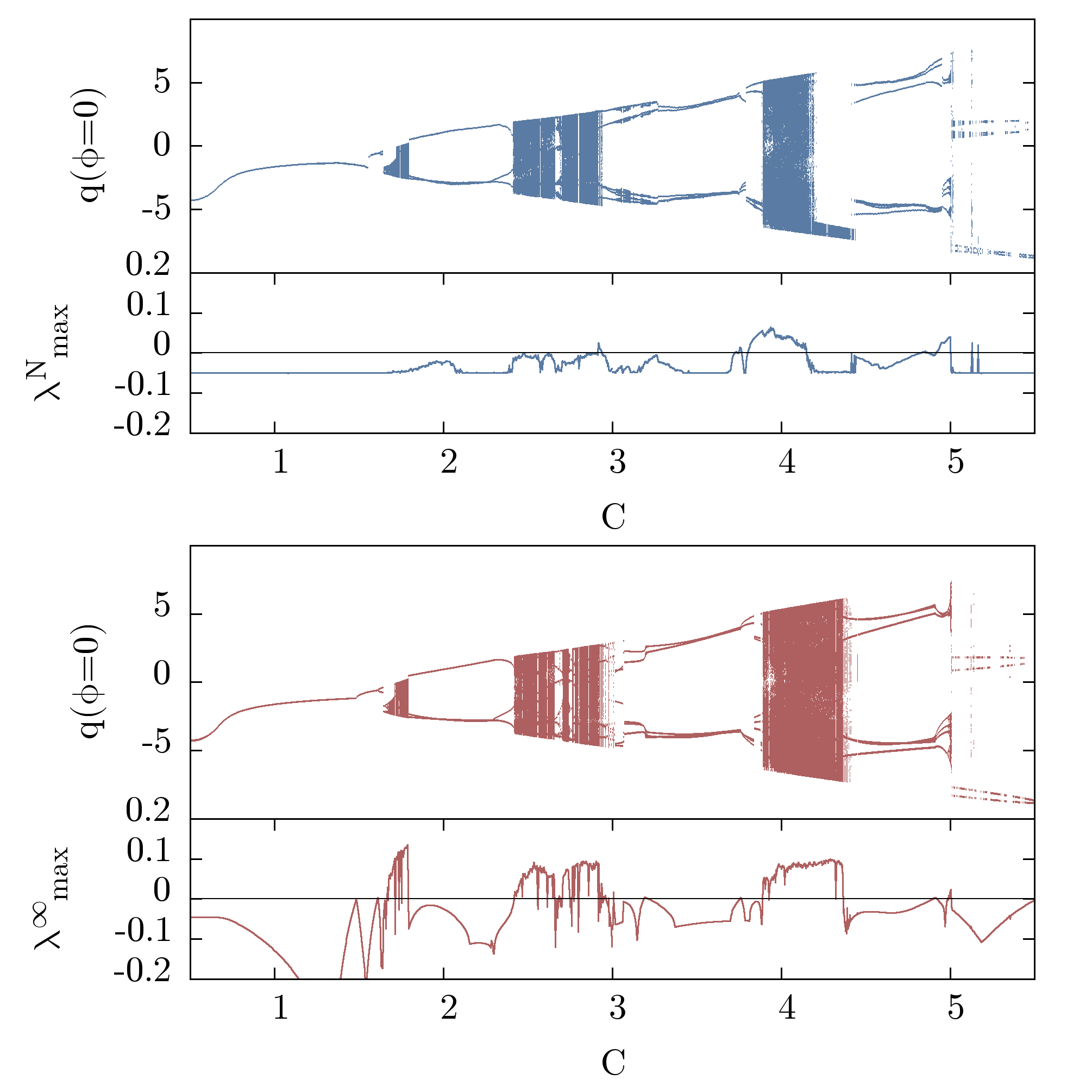}                         
    \caption{The bifurcation diagrams with the largest Lyapunov exponent for the piecewise-smooth system with a large number of spins $ N = 20\,000 $ ($\textcolor{blau}{\blacksquare}$, top) and for one special disorder disorder realization $ \{b_i\} $ does not show the typical bifurcation scenario, which is known from piecewise-smooth systems with grazing behavior. Instead, the behavior is very similar to the behavior of the continuous system ($\textcolor{rot}{\blacksquare}$, bottom) in the thermodynamic limit ($N = \infty$).}
    \label{fig:bif}
\end{figure}

\begin{figure*}
    \centering
    \includegraphics[width=1.0\textwidth]{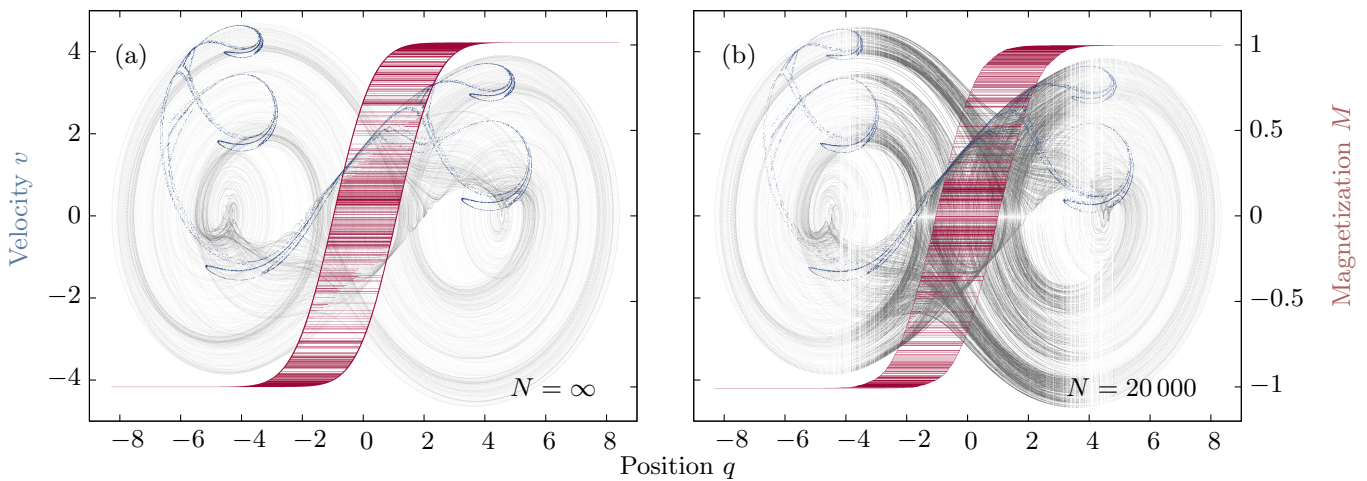}                         
    \caption{
    Comparison of the system in its thermodynamic limit (a) and the piecewise-smooth system with $N = 20\,000$ spins and for one specific disorder realization $ \{b_i\} $ (b). The light gray curves (\textcolor{grau}{\linie}) show the chaotic attractor, projected in the $q$-$v$-space and the red (dark gray) curves (\textcolor{rot}{\linie}) are the magnetization. For the system with $N=\infty$ the Poincar\'{e} section for $\phi=0$ is projected from the $q$-$v$-$w$-space into the $q$-$v$-space, whereas for $N=20\,000$ the Poincar\'{e} section is plotted in the $q$-$v$-space as a superposition of all $N/2+1$ sheets, illustrated by the blue (gray) points ($\textcolor{blau}{\bullet}$). For both systems we chose $C=4.0$.}    
    \label{fig:erf_10000}
\end{figure*}

For only a small number of spins the overall behavior of the piecewise-smooth system does not change much in comparison to the system with only two spins. An interesting questions arises, when the number of spins becomes very large ($N \to \infty$). On the one hand side, when the number of spins increases, also the number of boundaries increases. This means, when we determine the largest Lyapunov exponent, we have an increasing number of saltation matrices $\mt{X}$ multiplied with the fundamental solution $\mt{Y}$ (see Eq.\,\eqref{eq:salt_matrix} and \eqref{eq:lyap-ex}). On the other hand, the change in the magnetization, when crossing a single boundary, decreases and so the influence of each boundary becomes smaller because $ \mt{X} \to \mt{I} $. Hence $F_\text{hys}$ becomes smoother. In \cite{zech2020} we showed, that in this case chaos is still possible.

\textit{Many spins}: A comparison between the system in its thermodynamic limit ($N = \infty$) and the piecewise-smooth system for $ N = 20\,000$ spins is illustrated in Fig.\,\ref{fig:bif} and Fig.\,\ref{fig:erf_10000}, where the same parameters and initial conditions are used for both systems. From Fig.\,\ref{fig:erf_10000} it seems, that the trajectory, the Poincar\'{e} section and the magnetization curve behave in a similar way. In contrast the comparison of the bifurcation scenario for both systems ($\textcolor{blau}{\blacksquare}$,top $\mathrel{\hat=}$ piecewise-smooth; $\textcolor{rot}{\blacksquare}$,bottom $\mathrel{\hat=}$ thermodynamic limit), shows some differences (see Fig.\,\ref{fig:bif}). First of all, we find that the typical bifurcation scenarios for piecewise-smooth systems, which we found in the system with only a few spins, vanish. This is an indicator, that only transversal intersections of the trajectory with the boundaries are important for systems with a large number of spins. This is similar to the behavior we found in the system for independent spins \cite{zech2020}. In general, the two bifurcation diagrams show overall some similarity, but they differ within the periodic windows. This is due to the fact, that we only used one initial condition and always the same disorder realization. Like in the case of only one spin dimer, the exact position of the chaotic and periodic windows can change when changing these values. 

Another remarkable point is that the calculated largest Lyapunov exponent $\lambda_\text{max}^N$ of the piecewise-smooth system (determined by the theory of \ac{DM}, see Sec.\,\ref{sec:pws}) differs from $\lambda_\text{max}^\infty$ of the system in its thermodynamic limit (determined by extending the state variable by the relevant extreme values and linearizing the system, see Sec.\,\ref{sec:limit}). For a small coupling strength $ C \in [1.7,1.8] $ and $ C \in [2.5,3.0] $, by looking at the projection of the Poincar\'{e} section at the $q$-axes for $N=20\,000$, it seems, that the systems behaves in a chaotic way, which is in contrast to the finding of $\lambda_\text{max}^N < 0$ for these two windows. Therefore we suggest, that in contrast to the system in its thermodynamic limit with $\lambda_\text{max}^\infty > 0$, in this case the piecewise-smooth system does not show chaotic behavior, but undergoes a long periodic motion. In contrast, for larger values of $C$ (e.\,g. $ C \in [3.8,4.2] $) both systems show chaotic motion. As we mentioned at the beginning of this chapter, note, that for the piecewise-smooth system the Lyapunov spectrum is determined from a two-dimensional tangent space, whereas the Lyapunov spectrum for the system in its thermodynamic limit is calculated in a three-dimensional tangent space, neglecting the perturbation corresponding to the phase $\phi$ for both systems. It seems, that this difference leads to a different behavior of $\lambda_\text{max}$ for smaller values of the coupling strength $C$.

\subsection{Fractal dimension of the chaotic attractor}

For a quantitative measure of the similarity between the chaotic attractor of the piecewise-smooth system with a large but finite number of spins and the continuous system in the thermodynamic limit, we analyze the box counting $ D_\text{BC} $ and the Kaplan-Yorke dimension $ D_\text{KY} $ of the chaotic attractors of both systems. As we discussed before, $ D_\text{BC}^N $ is determined from the superposition of the $N/2+1$ sheets, whereas $ D_\text{BC}^\infty $ is determined from a projection into the $q$-$v$-space. Also $ D_\text{KY}^N $ is calculated by using two Lyapunov exponents and $ D_\text{KY}^\infty $ is, in general, calculated by using three Lyapunov exponents. Note, that $ D_\text{KY}^\infty $ is effectively determined by two Lyapunov exponents only, because of $ \lambda_4^\infty = -\infty $.

\begin{figure}
    \centering
    \includegraphics[width=1.0\columnwidth]{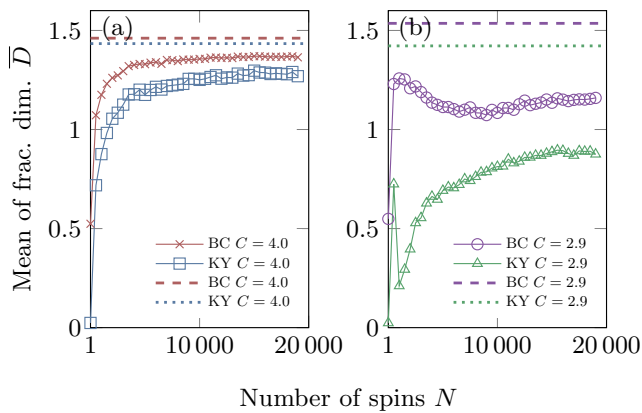}
    \caption{Calculation of the mean of the box counting (BC) and the Kaplan-Yorke dimension (KY) over $500$ disorder realizations for the piecewise-smooth system indicated by the different data point symbols. The dashed and dotted horizontal lines illustrate the corresponding fractal dimensions for the system in its thermodynamic limit. The coupling strength was chosen to $ C = 4.0 $ (a) and $ C = 2.9$ (b).}
    \label{fig:dim}
\end{figure}

We now focus on two examples with coupling strength $ C = 2.9 $ and $ C = 4.0 $. The results are shown in Fig.\,\ref{fig:dim}. On the one hand, we calculated the disorder average of both dimension $ \overline{D}_\text{BC} $ and $ \overline{D}_\text{KY} $ for the piecewise-smooth system with an increasing number of spins (solid lines), and on the other hand,  we calculated the same values in the thermodynamic limit $ N = \infty $ (dashed lines). For the piecewise-smooth system we take the average over $ 500 $ different realizations of the local disorder fields $b_i$ at each value of $N$. 

\begin{figure}
    \centering
    \includegraphics[width=1.0\columnwidth]{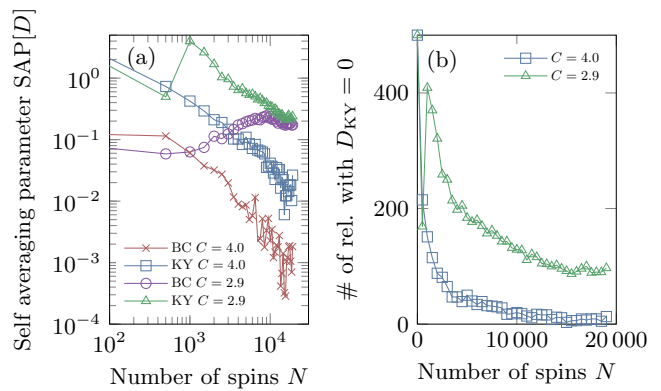}
    \caption{(a): For $ C = 4.0 $ and for $ C = 2.9 $ we determined the \ac{SAP} of the box counting and the Kaplan-Yorke dimension over $500$ disorder realizations for the piecewise-smooth system. The fractal dimensions shows self averaging behavior in case of $ C = 4.0 $. For $ C = 2.9 $ the box counting and the Kaplan-Yorke dimension do not show self averaging properties. (b): For $ C = 2.9 $ the number of realizations with $ D_\text{KY} = 0 $ converges to a constant value near $100$. One specific example is shown in Fig.\,\ref{fig:bif}\,(top). This indicates the non self-averaging property of the Kaplan-Yorke dimension.}
    \label{fig:sap}
\end{figure}

For $ C = 4.0 $ we can see, that both fractal dimensions of the piecewise-smooth system nearly converge to their values in the thermodynamic limit. This supports the assumption, that for larger values of $C$ and independent of the realization of the disorder $\{b_i\}$ of the piecewise-smooth system, both systems show chaotic motion. In contrast, for $ C = 2.9 $ the situation is not so clear. Even for $N=20\,000$ the averaged values for the fractal dimensions of the piecewise-smooth system are considerably different from the values in the thermodynamic limit. Remarkably, in the piecewise-smooth system the averaged Kaplan-Yorke dimension is smaller than one, which shows that for a nearly constant fraction of disorder realizations the system does not show chaos and $D_\text{KY}=0$ (see Fig.\,\ref{fig:sap}(b)). 

For a more detailed analysis in this direction, we calculated the coefficient of variation, also often called \ac{SAP} of the fractal dimensions, again for $ 500 $ different disorder realizations. The \ac{SAP} is given by:
\begin{equation}
    \operatorname{SAP}[D] = \frac{\overline{D^2} - \overline{D}^2}{\overline{D}^2} .
\end{equation}
In general, the \ac{SAP} specifies the relative variance of a quantity, here the fractal dimension. If it vanishes for $N \to \infty$, the behavior of the ensemble can be represented by only one realization of the local disorder with many spins. In Fig.\,\ref{fig:sap} the $ \operatorname{SAP}[D] $ is illustrated in dependency on the number of spins for $ C = 2.9 $ and $ C = 4.0 $. One can see, that for $ C = 4.0 $ we have $ \operatorname{SAP}[D] \to 0 $ for $ N \to \infty $, which supports the argument that the fractal dimension converges to the value in the thermodynamic limit. More precisely, the $ \operatorname{SAP}[D] $ converges to zero in an algebraic way, whereas in the case of independent spins we found that $ \operatorname{SAP}[D] $ converges to zero exponentially \cite{zech2020}. A different behavior can be observed for $ C = 2.9 $. In this case, a non-zero value remains for $\operatorname{SAP}[D]$, which means, that there is no self-averaging and there are at least two or more different representative values of the fractal dimensions for the ensemble. In fact, for $ C = 2.9 $ and a large number of spins nearly $100$ of the $500$ disorder realizations lead to $ D_\text{KY} = 0 $ (see Fig.\,\ref{fig:sap}\,(b)), whereas the remaining $400$ realizations are associated with a Kaplan-Yorke dimension larger than one. A specific example is that of Fig.\,\ref{fig:bif}(top) at $C=2.9$ and  $\lambda_\text{max}^N < 0$. This explains also the convergence of the averaged Kaplan-Yorke dimension $ \overline{D}_\text{KY} $ to a value between zero and one. Thus the thermodynamic limit $ N = \infty $ and the limit of the piecewise-smooth system with large but finite $ N $ are different for $ C = 2.9 $, which strongly differs from the behavior that we found for independent spins \cite{zech2020}.
The coexistence of regular and chaotic systems in the ensemble for $C=2.9$ explains the absence of self-averaging of the fractal dimension. This is supported by Fig.\,\ref{fig:sap}\,(b), which suggests, that a finite fraction of systems remains regular for $N \to \infty$.

\begin{figure}
    \centering
    \includegraphics[width=1.0\columnwidth]{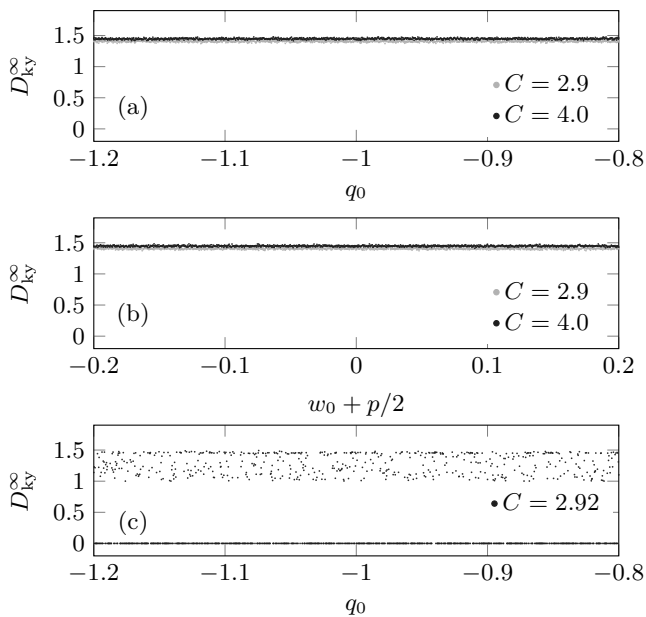}
    \caption{
    Kaplan-Yorke dimension of the system in its thermodynamic limit for three different values of $C$ in dependence on the initial values $q_0$ and $w_0$. The initial velocity was chosen as $v_0=0.1$. (a): The system starts outside the play for different values of $q_0$, hence $ w_0 = q_0 - p/2 $. For both values of $ C = 2.9 $ and $ C = 4.0 $ the Kaplan-Yorke dimension seems to be independent of the initial values of $q_0$. (b): The system starts inside the play for different values of $w_0$ and with $q_0=-1$. Again, for both values of $C$, $ D_\text{KY}^\infty $ seems to be independent of $w_0$. (c): In case of $C=2.92$ the system behaves totally different and ``switches'' between regular behavior with $ D_\text{KY}^\infty = 0 $ and chaotic motion with $ D_\text{KY}^\infty \in [1.0,1.5] $, respectively. More details are given in the text.
    }
    \label{fig:mboa}
\end{figure}

It is worth to mention, that for the chosen values of $ C = 2.9 $ and $ C = 4.0 $ the Kaplan-Yorke dimensions of the system in its thermodynamic limit $ D_\text{KY}^\infty $, illustrated by the dotted lines in Fig.\,\ref{fig:dim}\,(a) and (b), are independent on the initial values $q_0$ and $w_0$ of the system. To validate this, we first calculated $ D_\text{KY}^\infty $ for different values of $q_0$ (see Fig.\,\ref{fig:mboa}(a)), where the systems starts outside the play with $ w_0 = q_0 - p/2 $. In a second case, we determined $ D_\text{KY}^\infty $ for different values of $w_0$ (see Fig.\,\ref{fig:mboa}(b)), here the systems starts inside the play with $q_0=-1$. In both cases we chose $v_0 = 0.1$. For $ C = 2.9 $ and $ C = 4.0 $ we find the Kaplan-Yorke dimensions to stay within the limits $ D_\text{KY}^\infty \in [1.37,1.45] $ and $ D_\text{KY}^\infty \in [1.42,1.48] $, respectively. Hence for those values of $C$, we suggest, that for different initial conditions $q_0$ and $w_0$ the system shows chaotic motion. Because of the variation of $ D_\text{KY}^\infty $ we suggest, that in both cases the system reaches different attractors, but the corresponding Kaplan-Yorke dimensions seem to be very similar. For other values of $C$ this is not necessarily the case. For instance for $ C=2.92 $ (see Fig.\,\ref{fig:mboa}(c)) different initial values of $q_0$ can lead to regular behavior with $ D_\text{KY}^\infty = 0 $ or chaotic motion with $ D_\text{KY}^\infty \in [1.0,1.5] $, respectively. Thus, in the chaotic case the system again evolves to different attractors, but with totally different values of $ D_\text{KY}^\infty $. We also investigated the dependence on the initial conditions in the stable region for $ C = 1.9 $. We find, that in this case the system reaches different periodic orbits with slightly different negative Lyapunov exponents in dependence on $q_0$. This is a well know feature for dynamical systems coupled to hysteretic behavior \cite{radons2013nonlinear,radons2008dynamics}.

\subsection{Magnetization}

\begin{figure}
    \centering
    \includegraphics[width=1.0\columnwidth]{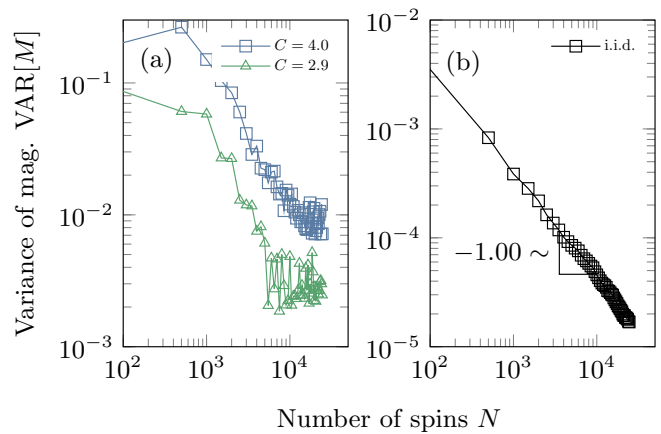}
    \caption{(a): Variance of the magnetization in dependency on the the number of spins of the piecewise-smooth system for $500$ disorder realizations. It can be seen, that for $ C = 2.9 $ the system does clearly not self-average with respect to the magnetization. Also for $ C = 4.0 $ it seems, that the system does not self-average. (b): Illustration of the variance of the magnetization for the system without dynamic feedback, but with iid input, here the variance of the the magnetization goes algebraically to zero as $ N^{-1} $.}
    \label{fig:mag}
\end{figure}

In addition to the fractal dimension we are interested in the behavior of the magnetization of the piecewise-smooth system in the context of its dynamics. Therefore we defined the time averaged magnetization $ M_T = \frac{1}{N} \sum_i \langle \sigma_i \rangle_\text{Time} $, where $ \langle \sigma_i \rangle $ denotes the time-average of the configuration of the $i$th spin, where, dependent on the dynamics of the system, $ \langle \sigma_i \rangle_\text{Time} $ does not necessarily equal zero. Here we simulated the system with a length of $\num{e6}$ time steps. Since, in general, the assemble average of the magnetization becomes zero $ \overline{M_T} = 0 $, we calculated the variance of the magnetization $ \operatorname{VAR}[M] $ instead of the \ac{SAP} of the magnetization. The variance is defined by
\begin{equation}
\operatorname{VAR}[M] = \overline{M_T^2} - \overline{M_T}^2 .
\end{equation}
Again the ensemble size for the quenched disorder is $500$ and we have chosen $ C = 2.9 $ and $ C = 4.0 $. The results are shown in Fig.\,\ref{fig:mag}.  
For $ C = 2.9 $ and also for $ C = 4.0 $ we find, that up to $ N \approx 5\,000 $ and $ N \approx 10\,000 $, respectively, the variance of both system decreases roughly algebraically with $ N^{-1} $. This behavior is equivalent to the behavior in a system without dynamic feedback but with independent and identically distributed external field input $B$, where the variance of the magnetization converges algebraically to zero in accordance with the central limit theorem. Instead for $ C = 2.9 $ in the coupled system, for a larger number of spins the decay of $ \operatorname{VAR}[M] $ becomes slower and eventually stays constant and non-zero for large $N$. This indicates, that the magnetization does not show self-averaging for $ C = 2.9 $. In case $ C = 4.0 $ it seems, that $ \operatorname{VAR}[M] $ also stays constant and non-zero for large $N$. From these findings, we can see, that the non self-averaging behavior of the magnetization is dynamically induced.

\begin{figure}
    \centering
    \includegraphics[width=1.0\columnwidth]{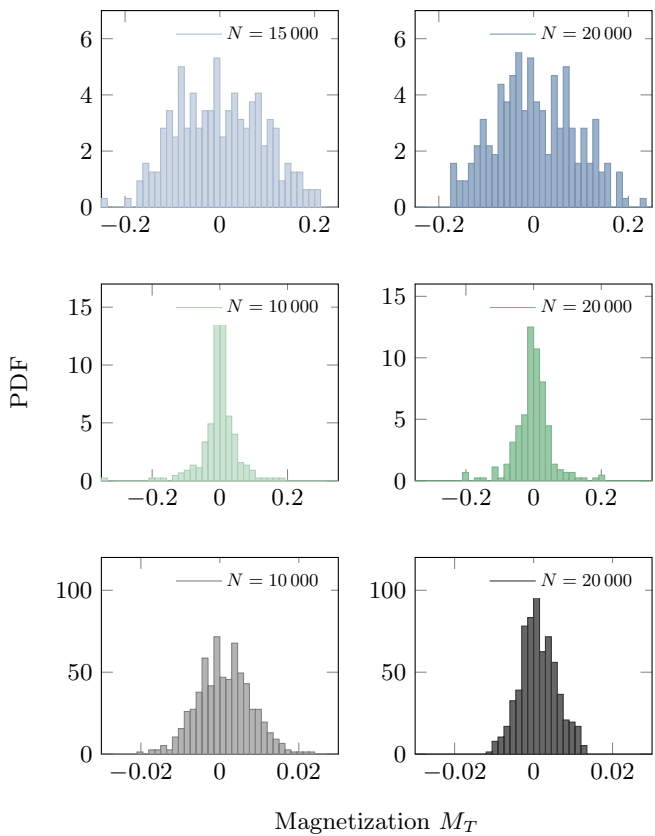}
    \caption{Histograms of the empirical \ac{PDF} for different values of $N$ in case of $ C = 4.0 $ (\textcolor{blau}{\linie}, top row), $ C = 2.9 $ (\textcolor{gruen}{\linie}, middle row) and for iid input (\textcolor{black}{\linie}, bottom row). In contrast to the magnetization for iid external field input $B$, in the system with feedback for both values of $ C $ the \ac{PDF} does not become narrower for increasing $ N $, which indicates the non self-averaging property of the magnetization.}
    \label{fig:mag_his}
\end{figure}

To verify the almost constant behavior of $ \operatorname{VAR}[M] $ for large $N$, we plotted the empirical \ac{PDF} of the time-averaged magnetization in Fig.\,\ref{fig:mag_his}. For $ C = 4.0 $ (\textcolor{blau}{\linie}, top row) and $ C = 2.9 $ (\textcolor{gruen}{\linie}, middle row), we show the \ac{PDF}, on the one hand, for $ N = 15\,000 $ and $N= 10\,000 $ spins, respectively, and on the other hand, for $ N = 20\,000 $ spins. In addition, the behavior of the \ac{PDF} for iid input is shown (\textcolor{black}{\linie}, bottom row). For the coupled systems and for $ C = 2.9 $ as well as for $ C = 4.0 $ the \ac{PDF} of the magnetization appears to become stationary, whereas for the iid input the variance of the distribution decreases for an increasing number of spins $N$. This means, that for a large number of spins different time-averaged magnetizations are possible for different realizations of the local disorder field, which we have found already in the system with independent spins \cite{zech2020}. 

\section{Conclusion} \label{sec:con}
In this paper we extended our results from the investigation of independent spins of a \ac{RFIM} coupled to a damped and periodically driven harmonic oscillator \cite{zech2020} to pairs of interacting spins. We showed, that two interacting spins can form an elementary hysteresis loop, depending on the local disorder fields of the spins and the strength of the nearest neighbor interaction. We analyzed a system with hysteretic play character by using only loops with the same width. We determined the relationship between an ensemble of independent spin dimers at zero temperature and the Preisach operator and calculated the related Preisach density for the system in its thermodynamic limit. We also introduced a new formalism to determine the whole Lyapunov spectrum of a dynamical system with a hysteretic nonlinearity in form of a generalized play operator.

From the numerically calculated bifurcation diagram for two spins, we showed that the behavior of few spins is very similar to the case of only one independent spin and typical bifurcation scenarios for piecewise-smooth systems can be found. For a larger number of spins we also calculated the bifurcation diagram of the projected Poincar\'{e}-section and the corresponding largest Lyapunov exponent. We stated that in case of the piecewise-smooth system, in general, there are three Lyapunov exponents according to the three-dimensional space $\{ \, q,v,\phi \, \mbox{mod}\,2\pi \, \}$, whereas for the system in its thermodynamic limit we have to consider an additional dimension $w$, caused by the hysteretic play. Eventually, for larger coupling strengths we found a good agreement between both systems, whereas for smaller values of $C$ the largest Lyapunov exponents differ.

For a more detailed explanation, we investigated the box counting and the Kaplan-Yorke dimension of the attractor for a small and a larger value of the coupling strength. We stated that for the piecewise-smooth and the continuous system on one hand side the Kaplan-Yorke dimension is calculated from three and four Lyapunov exponents, respectively and on the other hand the box counting dimension is determined from a superposition of $N/2+1$ sheets and a projection, respectively. Nevertheless we found, that at least in case of the larger coupling strength the fractal dimensions of the attractors of the piecewise-smooth system converge to their values of the system in its thermodynamic limit. But in contrast to the system with independent spins and no hysteretic behavior, here the self-averaging parameter of the fractal dimensions converge for increasing $N$ to zero in an algebraic way and not exponentially \cite{zech2020}. In contrast for a smaller coupling strength, it seems, that in the chaotic case, there is still a fraction of realizations of the piecewise-smooth system, where the system appears to behave in a regular way, even for a large number of spins.

Furthermore, we investigated the variance of the magnetization. We found, that in case of the smaller coupling strength the magnetization, as well as the fractal dimensions, does not show self-averaging. But also in case of the larger coupling strength the magnetization does not self-average, though the fractal dimensions does. This discrepancy between the self-averaging of the fractal dimension and the magnetization is not a contradiction, because, in general, the self-averaging property depends on the observable. The difference in the self-averaging behavior for different values of $C$ is due to the absence of a finite fraction of regular behavior in case $C=4.0$ and the existence of such a finite fraction for $C=2.9$. 

In general, we conclude, that for a system with memory in form of a hysteretic generalized play operator the limit $ N = \infty $ and $ N \to \infty $ are not necessarily equal, due to the dynamical feedback in the system. 

We mention two examples, where our findings could be relevant for systems with a hysteretic behavior in form of a generalized play. First, let's imagine the complex hysteretic behavior of a system is modeled continuously ($ N = \infty $) by using e.\,g. the generalized Prandtl-Ishlinskii model, which has the generalized play operator as its building blocks. This has been done before, e.\,g. in pseudoelastic shape memory alloys \cite{janaideh2009a,Shakiba2018,liang2020} and NiTi wires \cite{Yoong2021} or in modeling the state-of-charge of lithium-silicon cells \cite{Chayratsami2020}. By embedding such systems into a dynamical environment, one has to make sure, that the results, which are obtained from the continuous modeling can be applied to the real system. This is not necessarily the case, because the actual real system does not have to behave continuously, but rather in a discrete way ($ N \to \infty $), like magnetic materials (Barkhausen effect \cite{barkhausen1919}) or plastic deformation (plastic events \cite{Falk1998}). This leads to the second case, for which our findings could be relevant. Magnetic hysteresis in form of the generalized play represents the optimal situation of a fully remanent hysteresis loop, which is tried to be achieved e.\,g. with nanowires \cite{Spinu2004,Rotaru2011,DaCol2011}. Because of the discrete nature of the origin of the hysteresis, the ideal thermodynamic limit ($ N = \infty $) can not be approached. Hence, considering such system in a dynamical environment, quantities obtained from only one realization of the system, in general, does not represent the whole behavior of the system due to the non self-averaging property. This statement could also be extended to magnetic systems with arbitrary complex hysteresis by suggesting, that the differences we found for $ N = \infty $ and $ N \to \infty $ should remain, also in the case of arbitrary complex hysteresis.

Finally, our findings for the system in the thermodynamic limit ($N=\infty$) could be experimentally tested by a parallel connection of springs, dampers and a Prandtl element \cite{Prandtl1928}.

\section{Acknowledgements} \label{sec:ack}
We would like to thank Olav Hellwig for helpful discussions and valuable suggestions.

\appendix*
\label{sec:app}
\begin{center}
   \Large \textsc{Appendix: Lyapunov exponents in the thermodynamic limit}
\end{center}

\noindent Here we want to determine the Lyapunov exponents of the damped harmonic oscillator with periodic forcing and an additional external hysteretic force in form of a generalized play operator. The equation of motion is given by Eq.\,\eqref{eq:eom_2}:
\begin{equation}
\dot{\vc{x}}(t)=
     \begin{pmatrix}
                v(t) \\
                 - 2\zeta v(t) - q(t) + \cos \phi(t) + C M[q](t) \\
                \Omega
                \end{pmatrix} , \label{eq:001}
\end{equation}
where $ M[q](t) $ denotes the hysteretic force given by Eq.\,\eqref{eq:prob_mag}:
\begin{equation}
M[q](t) = 2  \int\limits_{-\infty}^{w(t)} \mu(s) \, \mathrm{d}s - 1 = \operatorname{erf}\left( \frac{w(t)}{\sqrt{2}R} \right) . \label{eq:002}
\end{equation}
Here $ \mu(s) $ is the density of the center of the elementary hysteresis loops, which in our case is chosen to be Gaussian distributed. Thus the integration over $ \mu(s) $ leads to an error function in Eq.\,\eqref{eq:002}. Furthermore $ w(t) $ denotes the output of the play operator, which can be imagined as the center of the wagon in Fig.\,\ref{fig:play_mech}(a), and is given by Eq.\,\eqref{eq:real_play}:
\begin{equation}
    w(t) = \operatorname{max} \biggl\{ q(t)-\frac{p}{2} ,\operatorname{min} \biggl\{ q(t) + \frac{p}{2}, w(t_i) \biggr\} \biggr\} ,
    \label{eq:003}
\end{equation}
where $ t_i < t $ is the time of the last extremum of the input $ q(t') $ with $ t' < t $. Because of this, $ w(t) $ stores the position $ w(t_i) $, where the system enters into the play region. This expresses the presence of memory in the system. Hence $ w(t) $ is not a function of time, but a functional of the whole trajectory of the input $ q(t) $. In the following we want to take this into account by writing $ w(t) = \mathcal{W}[q](t) $. Now Eq.\,\eqref{eq:001} can be written as a functional differential equation
\begin{equation}
    \dot{\vc{x}}(t) = \vc{f}(\vc{x},w(t)) = \vc{f}(\vc{x},\mathcal{W}[q](t)) . \label{eq:004} 
\end{equation}
Formally, because of the functional character, this equation is infinite dimensional. But we will show that this reduces to four dimensions. To see this, note that instead of monitoring the evolution of $\vc{x}(t)$ as a solution of Eq.\,\eqref{eq:004}, it is convenient to monitor the evolution of $ w(t) $ separately and therefore consider the evolution of an instantaneous extended state variable $ \vc{y}(t) = (\vc{x}(t),w(t))^T$. Also note, that we have to choose the initial condition $\vc{y}(0) = (\vc{x}(0),w(0))^T$ in a special way, so that the constraint $ |q(0)-w(0)| < p/2 $ is full filled. 

To calculate the Lyapunov exponents, we have to linearize Eq.\,\eqref{eq:004} around a reference solution $\vc{x}^R(t) = \vc{x}(t) - \del \vc{x}(t)$. Because the perturbation $ \del \vc{x}(t) $ leads to a change in $ w(t) = w^R(t) + \del w(t) $, with
\begin{equation}
    \del w(t) = w(t) - w^R(t) = \mathcal{W}[q^R + \del q](t) - \mathcal{W}[q^R](t)
\end{equation}
we get:
\begin{align}
    \dot{\vc{x}}^R(t) + \del \dot{\vc{x}}(t) &= \vc{f}(\vc{x}^R(t) + \del \vc{x}(t),\mathcal{W}[q^R + \del q](t)) \notag \\
    \dot{\vc{x}}^R(t) + \del \dot{\vc{x}}(t) &= \vc{f}(\vc{x}^R(t) + \del \vc{x}(t),\mathcal{W}[q^R](t) + \del w(t)) . \label{eq:005}
\end{align}
By Taylor expanding this up to first order in $ \del \vc{x}(t) $ and $ \del w(t) $ it follows:
\begin{equation}
    \del \dot{\vc{x}}(t) = \underbrace{\left. \frac{\pd \vc{f}(\vc{\xi},\mathcal{W}[q^R](t))}{\pd \vc{\xi}} \right|_{\vc{x}^R(t)}}_{=: \mt{A}(t)} \del \vc{x} + \underbrace{\left. \frac{\pd \vc{f}(\vc{x}^R(t),\xi)}{\pd \xi} \right|_{w^R(t)}}_{=: \vc{B}(t)} \del w(t) . \label{eq:006}
\end{equation} 
With Eq.\,\eqref{eq:001} for the matrix $\mt{A}(t)$ we find:
\begin{equation}
    \mt{A}(t) = \begin{pmatrix}
                     0 & 1 & 0 \\
                    -1 & - 2 \zeta & - \sin \, \phi^R(t) \\
                    0 & 0 & 0
                \end{pmatrix} . \label{eq:007}
\end{equation}

For the following calculations it is appropriate to separate the behavior of $w(t)$ into two different cases (see Fig.\,\ref{fig:play_mech}(a)). We want to call the case, when the finger ($q^R$) is inside the wagon \textit{inside} and the case, where the finger touches one of the walls of the wagon \textit{outside}. We are taking this into account, by using the indicator function $ \chi^R(t) $ given in Eq.\,\eqref{eq:indicator}:
\begin{equation}
    \chi^R(t) = \begin{cases}
            0,  & \text{for } |w^R(t) - q^R(t)| < p/2, \text{ inside} , \\
            1,  & \text{for } |w^R(t) - q^R(t)| = p/2, \text{ outside} .
            \end{cases} \label{eq:008}
\end{equation}

Now we want to determine the term $ \del w(t) = \del \mathcal{W}[q](t) $ in Eq.\,\eqref{eq:006}. In general this differential of the functional $ \mathcal{W}[q](t) $ is given by:
\begin{equation}
    \del w(t) = \del \mathcal{W}[q](t) = \int\limits_0^t \frac{\del w(t)}{\del q(t')} \del q(t') \, \tpd t' . \label{eq:011}
\end{equation}
Again, if we are inside the play ($\chi^R=0$), here $ w(t) = w(t_i) $ becomes constant. Thus we have $ \frac{\del w(t)}{\del q(t')} = \frac{\del w(t_i)}{\del q(t')} = 0 $, except if the last extremum of $q(t)$ at $t=t_i$ also determined the value $w(t_i)$. Denoting the time of such an extremum as $t_i'$, we have $w(t)=q(t_i') \pm p/2$, thus $\frac{\del w(t)}{\del q(t')}=\frac{\del q(t_i') \pm p/2}{\del q(t')} = \del(t_i' - t')$. In the other case, if we are outside of the play ($\chi^R=1$), we have $ w(t) = q(t) \pm p/2 $, which gives us $\frac{\del w(t)}{\del q(t')} = \frac{\del q(t) \pm p/2}{\del q(t')} = \del (t-t')$. Inserting this into Eq.\,\eqref{eq:011} leads to:
\begin{equation}
    \del w(t) = \chi^R(t) \del q(t) + (1-\chi^R(t)) \del q(t_i') . \label{eq:012}
\end{equation}
Taking the time derivative gives us:
\begin{equation}
    \del \dot{w}(t) = \chi^R(t) \del \dot{q}(t) = \chi^R(t) \del v(t) . \label{eq:013}
\end{equation}

Furthermore we can calculate the vector $ \mt{B}(t) $ in Eq.\,\eqref{eq:006}. In case we are inside the play ($\chi^R=0$) the term $ M[q](t) $ in Eq.\,\eqref{eq:001} becomes constant with $ M[q](t) = \operatorname{erf}( w(t_i')/\sqrt{2}R ) $ and we find:
\begin{equation}
    \vc{B}^0(t_i') = \begin{pmatrix}
                     0  \\
                    \left. C \frac{\pd M(\xi)}{\pd \xi} \right|_{w^R(t_i')} \\
                    0 
                \end{pmatrix}  . \label{eq:009}
\end{equation}
In contrast, if we are outside of the play ($\chi^R=1$) we have $ M[q](t) = \operatorname{erf}( w(t)/\sqrt{2}R ) $ and therefore:
\begin{equation}
    \vc{B}^1(t) = \begin{pmatrix}
                     0  \\
                    \left. C \frac{\pd M(\xi)}{\pd \xi} \right|_{w^R(t)} \\
                    0 
                \end{pmatrix}  . \label{eq:010}
\end{equation}

Putting it all together and by using the extended state variable $\del \vc{y} = (\del q,\del v,\del \phi,\del w)^T$ we can finally write down the full system of differential equations for the perturbations: \\ \noindent \textbf{Inside: $\chi^R=0$}
\begin{equation}
    \del \dot{\vc{y}}(t) = 
    \underbrace{\begin{pmatrix}
      0 & 1 & 0 & 0 \\
      -1 & - 2 \zeta & - \sin \, \phi^R(t) & \left. C \frac{\pd M(\xi)}{\pd \xi} \right|_{w^R(t_i')} \\
       0 & 0 & 0 & 0 \\
       0 & 0 & 0 & 0
    \end{pmatrix}}_{=:\mt{W}^0(t,t_i')}
    \del \vc{y}(t)  .
\end{equation}

~\\ \noindent \textbf{Outside: $\chi^R=1$}
\begin{equation}
    \del \dot{\vc{y}}(t) = 
    \underbrace{\begin{pmatrix}
      0 & 1 & 0 & 0 \\
      -1 & - 2 \zeta & - \sin \, \phi^R(t) & \left. C \frac{\pd M(\xi)}{\pd \xi} \right|_{w^R(t)} \\
       0 & 0 & 0 & 0 \\
       0 & 1 & 0 & 0
    \end{pmatrix}}_{=:\mt{W}^1(t)}
    \del \vc{y}(t) .
\end{equation}


\begin{acronym}
\acro{PM}[PM]{Preisach Model}
\acro{RFIM}[RFIM]{Random Field Ising Model} 
\acro{ODE}[ODE]{Ordinary Differential Equation} 
\acro{EOM}[EOM]{Equation Of Motion}
\acro{DM}[DM]{Discontinuity Mapping}
\acro{ZDM}[ZDM]{Zero Time Discontinuity Mapping}
\acro{PDM}[PDM]{Poincar\'{e}-section Discontinuity Mapping}
\acro{SAP}[SAP]{Self-averaging Parameter}
\acro{iid}[iid]{Independent and Identically Distributed}
\acro{PDF}[PDF]{Probability Density Function}
\end{acronym}
    
\bibliography{main}

\begin{thebibliography}{85}%
\makeatletter
\providecommand \@ifxundefined [1]{%
 \@ifx{#1\undefined}
}%
\providecommand \@ifnum [1]{%
 \ifnum #1\expandafter \@firstoftwo
 \else \expandafter \@secondoftwo
 \fi
}%
\providecommand \@ifx [1]{%
 \ifx #1\expandafter \@firstoftwo
 \else \expandafter \@secondoftwo
 \fi
}%
\providecommand \natexlab [1]{#1}%
\providecommand \enquote  [1]{``#1''}%
\providecommand \bibnamefont  [1]{#1}%
\providecommand \bibfnamefont [1]{#1}%
\providecommand \citenamefont [1]{#1}%
\providecommand \href@noop [0]{\@secondoftwo}%
\providecommand \href [0]{\begingroup \@sanitize@url \@href}%
\providecommand \@href[1]{\@@startlink{#1}\@@href}%
\providecommand \@@href[1]{\endgroup#1\@@endlink}%
\providecommand \@sanitize@url [0]{\catcode `\\12\catcode `\$12\catcode
  `\&12\catcode `\#12\catcode `\^12\catcode `\_12\catcode `\%12\relax}%
\providecommand \@@startlink[1]{}%
\providecommand \@@endlink[0]{}%
\providecommand \url  [0]{\begingroup\@sanitize@url \@url }%
\providecommand \@url [1]{\endgroup\@href {#1}{\urlprefix }}%
\providecommand \urlprefix  [0]{URL }%
\providecommand \Eprint [0]{\href }%
\providecommand \doibase [0]{https://doi.org/}%
\providecommand \selectlanguage [0]{\@gobble}%
\providecommand \bibinfo  [0]{\@secondoftwo}%
\providecommand \bibfield  [0]{\@secondoftwo}%
\providecommand \translation [1]{[#1]}%
\providecommand \BibitemOpen [0]{}%
\providecommand \bibitemStop [0]{}%
\providecommand \bibitemNoStop [0]{.\EOS\space}%
\providecommand \EOS [0]{\spacefactor3000\relax}%
\providecommand \BibitemShut  [1]{\csname bibitem#1\endcsname}%
\let\auto@bib@innerbib\@empty
\bibitem [{\citenamefont {Gusev}\ \emph {et~al.}(2020)\citenamefont {Gusev},
  \citenamefont {Sadovnikov}, \citenamefont {Nikitov}, \citenamefont
  {Sapozhnikov},\ and\ \citenamefont {Udalov}}]{Gusev2020}%
  \BibitemOpen
  \bibfield  {author} {\bibinfo {author} {\bibfnamefont {N.~S.}\ \bibnamefont
  {Gusev}}, \bibinfo {author} {\bibfnamefont {A.~V.}\ \bibnamefont
  {Sadovnikov}}, \bibinfo {author} {\bibfnamefont {S.~A.}\ \bibnamefont
  {Nikitov}}, \bibinfo {author} {\bibfnamefont {M.~V.}\ \bibnamefont
  {Sapozhnikov}},\ and\ \bibinfo {author} {\bibfnamefont {O.~G.}\ \bibnamefont
  {Udalov}},\ }\bibfield  {title} {\bibinfo {title} {Manipulation of the
  {D}zyaloshinskii--{Mo}riya interaction in co/pt multilayers with strain},\
  }\href@noop {} {\bibfield  {journal} {\bibinfo  {journal} {Phys. Rev. Lett.}\
  }\textbf {\bibinfo {volume} {124}},\ \bibinfo {pages} {157202} (\bibinfo
  {year} {2020})}\BibitemShut {NoStop}%
\bibitem [{\citenamefont {Vlasko-Vlasov}\ \emph {et~al.}(2020)\citenamefont
  {Vlasko-Vlasov}, \citenamefont {Welp}, \citenamefont {Koshelev},
  \citenamefont {Smylie}, \citenamefont {Bao}, \citenamefont {Chung},
  \citenamefont {Kanatzidis},\ and\ \citenamefont {Kwok}}]{Vlasko-Vlasov2020}%
  \BibitemOpen
  \bibfield  {author} {\bibinfo {author} {\bibfnamefont {V.~K.}\ \bibnamefont
  {Vlasko-Vlasov}}, \bibinfo {author} {\bibfnamefont {U.}~\bibnamefont {Welp}},
  \bibinfo {author} {\bibfnamefont {A.~E.}\ \bibnamefont {Koshelev}}, \bibinfo
  {author} {\bibfnamefont {M.}~\bibnamefont {Smylie}}, \bibinfo {author}
  {\bibfnamefont {J.-K.}\ \bibnamefont {Bao}}, \bibinfo {author} {\bibfnamefont
  {D.~Y.}\ \bibnamefont {Chung}}, \bibinfo {author} {\bibfnamefont {M.~G.}\
  \bibnamefont {Kanatzidis}},\ and\ \bibinfo {author} {\bibfnamefont {W.-K.}\
  \bibnamefont {Kwok}},\ }\bibfield  {title} {\bibinfo {title} {Cooperative
  response of magnetism and superconductivity in the magnetic superconductor
  $\mathrm{RbEuF}{\mathrm{e}}_{4}\mathrm{A}{\mathrm{s}}_{4}$},\ }\href@noop {}
  {\bibfield  {journal} {\bibinfo  {journal} {Phys. Rev. B}\ }\textbf {\bibinfo
  {volume} {101}},\ \bibinfo {pages} {104504} (\bibinfo {year}
  {2020})}\BibitemShut {NoStop}%
\bibitem [{\citenamefont {Moorsom}\ \emph {et~al.}(2020)\citenamefont
  {Moorsom}, \citenamefont {Alghamdi}, \citenamefont {Stansill}, \citenamefont
  {Poli}, \citenamefont {Teobaldi}, \citenamefont {Beg}, \citenamefont
  {Fangohr}, \citenamefont {Rogers}, \citenamefont {Aslam}, \citenamefont
  {Ali}, \citenamefont {Hickey},\ and\ \citenamefont {Cespedes}}]{Moorsom2020}%
  \BibitemOpen
  \bibfield  {author} {\bibinfo {author} {\bibfnamefont {T.}~\bibnamefont
  {Moorsom}}, \bibinfo {author} {\bibfnamefont {S.}~\bibnamefont {Alghamdi}},
  \bibinfo {author} {\bibfnamefont {S.}~\bibnamefont {Stansill}}, \bibinfo
  {author} {\bibfnamefont {E.}~\bibnamefont {Poli}}, \bibinfo {author}
  {\bibfnamefont {G.}~\bibnamefont {Teobaldi}}, \bibinfo {author}
  {\bibfnamefont {M.}~\bibnamefont {Beg}}, \bibinfo {author} {\bibfnamefont
  {H.}~\bibnamefont {Fangohr}}, \bibinfo {author} {\bibfnamefont
  {M.}~\bibnamefont {Rogers}}, \bibinfo {author} {\bibfnamefont
  {Z.}~\bibnamefont {Aslam}}, \bibinfo {author} {\bibfnamefont
  {M.}~\bibnamefont {Ali}}, \bibinfo {author} {\bibfnamefont {B.~J.}\
  \bibnamefont {Hickey}},\ and\ \bibinfo {author} {\bibfnamefont
  {O.}~\bibnamefont {Cespedes}},\ }\bibfield  {title} {\bibinfo {title}
  {$\ensuremath{\pi}$-anisotropy: A nanocarbon route to hard magnetism},\
  }\href@noop {} {\bibfield  {journal} {\bibinfo  {journal} {Phys. Rev. B}\
  }\textbf {\bibinfo {volume} {101}},\ \bibinfo {pages} {060408} (\bibinfo
  {year} {2020})}\BibitemShut {NoStop}%
\bibitem [{\citenamefont {Das}\ \emph {et~al.}(2021)\citenamefont {Das},
  \citenamefont {Krug},\ and\ \citenamefont {Mungan}}]{das2021}%
  \BibitemOpen
  \bibfield  {author} {\bibinfo {author} {\bibfnamefont {S.~G.}\ \bibnamefont
  {Das}}, \bibinfo {author} {\bibfnamefont {J.}~\bibnamefont {Krug}},\ and\
  \bibinfo {author} {\bibfnamefont {M.}~\bibnamefont {Mungan}},\ }\href@noop {}
  {\bibinfo {title} {A driven disordered systems approach to biological
  evolution in changing environments}} (\bibinfo {year} {2021}),\ \Eprint
  {https://arxiv.org/abs/2108.06170} {arXiv:2108.06170 [q-bio.PE]} \BibitemShut
  {NoStop}%
\bibitem [{\citenamefont {Bertotti}\ and\ \citenamefont
  {Mayergoyz}(2006)}]{Bertotti2006}%
  \BibitemOpen
  \bibfield  {author} {\bibinfo {author} {\bibfnamefont {G.}~\bibnamefont
  {Bertotti}}\ and\ \bibinfo {author} {\bibfnamefont {I.~D.}\ \bibnamefont
  {Mayergoyz}},\ }\href@noop {} {\emph {\bibinfo {title} {The science of
  hysteresis}}},\ Vol.\ \bibinfo {volume} {1-3}\ (\bibinfo  {publisher}
  {Academic Press},\ \bibinfo {address} {New York},\ \bibinfo {year}
  {2006})\BibitemShut {NoStop}%
\bibitem [{\citenamefont {Radons}(2008{\natexlab{a}})}]{radons2008hysteresis}%
  \BibitemOpen
  \bibfield  {author} {\bibinfo {author} {\bibfnamefont {G.}~\bibnamefont
  {Radons}},\ }\bibfield  {title} {\bibinfo {title} {Hysteresis-induced
  long-time tails},\ }\href@noop {} {\bibfield  {journal} {\bibinfo  {journal}
  {Phys. Rev. Lett.}\ }\textbf {\bibinfo {volume} {100}},\ \bibinfo {pages}
  {240602/1} (\bibinfo {year} {2008}{\natexlab{a}})}\BibitemShut {NoStop}%
\bibitem [{\citenamefont {Radons}(2008{\natexlab{b}})}]{radons2008spectral1}%
  \BibitemOpen
  \bibfield  {author} {\bibinfo {author} {\bibfnamefont {G.}~\bibnamefont
  {Radons}},\ }\bibfield  {title} {\bibinfo {title} {Spectral properties of the
  {P}reisach hysteresis model with random input. i. general results},\
  }\href@noop {} {\bibfield  {journal} {\bibinfo  {journal} {Phys. Rev. E}\
  }\textbf {\bibinfo {volume} {77}},\ \bibinfo {pages} {061133/1} (\bibinfo
  {year} {2008}{\natexlab{b}})}\BibitemShut {NoStop}%
\bibitem [{\citenamefont {Radons}(2008{\natexlab{c}})}]{radons2008spectral2}%
  \BibitemOpen
  \bibfield  {author} {\bibinfo {author} {\bibfnamefont {G.}~\bibnamefont
  {Radons}},\ }\bibfield  {title} {\bibinfo {title} {Spectral properties of the
  {P}reisach hysteresis model with random input. ii. universality classes for
  symmetric elementary loops},\ }\href@noop {} {\bibfield  {journal} {\bibinfo
  {journal} {Phys. Rev. E}\ }\textbf {\bibinfo {volume} {77}},\ \bibinfo
  {pages} {061134/1} (\bibinfo {year} {2008}{\natexlab{c}})}\BibitemShut
  {NoStop}%
\bibitem [{\citenamefont {Jiang}\ \emph {et~al.}(2010)\citenamefont {Jiang},
  \citenamefont {Ji}, \citenamefont {Qiu},\ and\ \citenamefont
  {Chen}}]{Jiang2010}%
  \BibitemOpen
  \bibfield  {author} {\bibinfo {author} {\bibfnamefont {H.}~\bibnamefont
  {Jiang}}, \bibinfo {author} {\bibfnamefont {H.}~\bibnamefont {Ji}}, \bibinfo
  {author} {\bibfnamefont {J.}~\bibnamefont {Qiu}},\ and\ \bibinfo {author}
  {\bibfnamefont {Y.}~\bibnamefont {Chen}},\ }\bibfield  {title} {\bibinfo
  {title} {A modified {P}randtl-{I}shlinskii model for modeling asymmetric
  hysteresis of piezoelectric actuators},\ }\href@noop {} {\bibfield  {journal}
  {\bibinfo  {journal} {IEEE Transactions on Ultrasonics, Ferroelectrics, and
  Frequency Control}\ }\textbf {\bibinfo {volume} {57}},\ \bibinfo {pages}
  {1200} (\bibinfo {year} {2010})}\BibitemShut {NoStop}%
\bibitem [{\citenamefont {Schubert}\ and\ \citenamefont
  {Radons}(2017)}]{schubert2017}%
  \BibitemOpen
  \bibfield  {author} {\bibinfo {author} {\bibfnamefont {S.}~\bibnamefont
  {Schubert}}\ and\ \bibinfo {author} {\bibfnamefont {G.}~\bibnamefont
  {Radons}},\ }\bibfield  {title} {\bibinfo {title} {{P}reisach models of
  hysteresis driven by markovian input processes},\ }\href@noop {} {\bibfield
  {journal} {\bibinfo  {journal} {Phys. Rev. E}\ }\textbf {\bibinfo {volume}
  {96}},\ \bibinfo {pages} {022117} (\bibinfo {year} {2017})}\BibitemShut
  {NoStop}%
\bibitem [{\citenamefont {Urbanaviciute}\ \emph {et~al.}(2018)\citenamefont
  {Urbanaviciute}, \citenamefont {Cornelissen}, \citenamefont {Meng},
  \citenamefont {Sijbesma},\ and\ \citenamefont
  {Kemerink}}]{urbanaviciute2018}%
  \BibitemOpen
  \bibfield  {author} {\bibinfo {author} {\bibfnamefont {I.}~\bibnamefont
  {Urbanaviciute}}, \bibinfo {author} {\bibfnamefont {T.~D.}\ \bibnamefont
  {Cornelissen}}, \bibinfo {author} {\bibfnamefont {X.}~\bibnamefont {Meng}},
  \bibinfo {author} {\bibfnamefont {R.~P.}\ \bibnamefont {Sijbesma}},\ and\
  \bibinfo {author} {\bibfnamefont {M.}~\bibnamefont {Kemerink}},\ }\bibfield
  {title} {\bibinfo {title} {Physical reality of the {P}reisach model for
  organic ferroelectrics},\ }\href@noop {} {\bibfield  {journal} {\bibinfo
  {journal} {Nature Communications}\ }\textbf {\bibinfo {volume} {9}},\
  \bibinfo {pages} {4409} (\bibinfo {year} {2018})}\BibitemShut {NoStop}%
\bibitem [{\citenamefont {Sethna}\ \emph {et~al.}(1993)\citenamefont {Sethna},
  \citenamefont {Dahmen}, \citenamefont {Kartha}, \citenamefont {Krumhansl},
  \citenamefont {Roberts},\ and\ \citenamefont {Shore}}]{Sethna1993}%
  \BibitemOpen
  \bibfield  {author} {\bibinfo {author} {\bibfnamefont {J.~P.}\ \bibnamefont
  {Sethna}}, \bibinfo {author} {\bibfnamefont {K.}~\bibnamefont {Dahmen}},
  \bibinfo {author} {\bibfnamefont {S.}~\bibnamefont {Kartha}}, \bibinfo
  {author} {\bibfnamefont {J.~A.}\ \bibnamefont {Krumhansl}}, \bibinfo {author}
  {\bibfnamefont {B.~W.}\ \bibnamefont {Roberts}},\ and\ \bibinfo {author}
  {\bibfnamefont {J.~D.}\ \bibnamefont {Shore}},\ }\bibfield  {title} {\bibinfo
  {title} {Hysteresis and hierarchies: Dynamics of disorder-driven first-order
  phase transformations},\ }\href@noop {} {\bibfield  {journal} {\bibinfo
  {journal} {Phys. Rev. Lett.}\ }\textbf {\bibinfo {volume} {70}},\ \bibinfo
  {pages} {3347} (\bibinfo {year} {1993})}\BibitemShut {NoStop}%
\bibitem [{\citenamefont {Lilly}\ \emph {et~al.}(1996)\citenamefont {Lilly},
  \citenamefont {Wootters},\ and\ \citenamefont {Hallock}}]{Lilly1996}%
  \BibitemOpen
  \bibfield  {author} {\bibinfo {author} {\bibfnamefont {M.~P.}\ \bibnamefont
  {Lilly}}, \bibinfo {author} {\bibfnamefont {A.~H.}\ \bibnamefont
  {Wootters}},\ and\ \bibinfo {author} {\bibfnamefont {R.~B.}\ \bibnamefont
  {Hallock}},\ }\bibfield  {title} {\bibinfo {title} {Spatially extended
  avalanches in a hysteretic capillary condensation system: Superfluid he-4 in
  nuclepore},\ }\href@noop {} {\bibfield  {journal} {\bibinfo  {journal} {Phys.
  Rev. Lett.}\ }\textbf {\bibinfo {volume} {77}},\ \bibinfo {pages} {4222}
  (\bibinfo {year} {1996})}\BibitemShut {NoStop}%
\bibitem [{\citenamefont {Shukla}(2000)}]{Shukla2000}%
  \BibitemOpen
  \bibfield  {author} {\bibinfo {author} {\bibfnamefont {P.}~\bibnamefont
  {Shukla}},\ }\bibfield  {title} {\bibinfo {title} {Exact solution of return
  hysteresis loops in a one-dimensional random-field {I}sing model at zero
  temperature},\ }\href@noop {} {\bibfield  {journal} {\bibinfo  {journal}
  {Phys. Rev. E}\ }\textbf {\bibinfo {volume} {62}},\ \bibinfo {pages} {4725}
  (\bibinfo {year} {2000})}\BibitemShut {NoStop}%
\bibitem [{\citenamefont {Sabhapandit}\ \emph {et~al.}(2002)\citenamefont
  {Sabhapandit}, \citenamefont {Dhar},\ and\ \citenamefont
  {Shukla}}]{Sabhapandit2002}%
  \BibitemOpen
  \bibfield  {author} {\bibinfo {author} {\bibfnamefont {S.}~\bibnamefont
  {Sabhapandit}}, \bibinfo {author} {\bibfnamefont {D.}~\bibnamefont {Dhar}},\
  and\ \bibinfo {author} {\bibfnamefont {P.}~\bibnamefont {Shukla}},\
  }\bibfield  {title} {\bibinfo {title} {Hysteresis in the random-field {I}sing
  model and bootstrap percolation},\ }\href@noop {} {\bibfield  {journal}
  {\bibinfo  {journal} {Phys. Rev. Lett.}\ }\textbf {\bibinfo {volume} {88}},\
  \bibinfo {pages} {197202} (\bibinfo {year} {2002})}\BibitemShut {NoStop}%
\bibitem [{\citenamefont {Deutsch}\ \emph {et~al.}(2004)\citenamefont
  {Deutsch}, \citenamefont {Dhar},\ and\ \citenamefont
  {Narayan}}]{Deutsch2004}%
  \BibitemOpen
  \bibfield  {author} {\bibinfo {author} {\bibfnamefont {J.~M.}\ \bibnamefont
  {Deutsch}}, \bibinfo {author} {\bibfnamefont {A.}~\bibnamefont {Dhar}},\ and\
  \bibinfo {author} {\bibfnamefont {O.}~\bibnamefont {Narayan}},\ }\bibfield
  {title} {\bibinfo {title} {Return to return point memory},\ }\href@noop {}
  {\bibfield  {journal} {\bibinfo  {journal} {Phys. Rev. Lett.}\ }\textbf
  {\bibinfo {volume} {92}},\ \bibinfo {pages} {227203} (\bibinfo {year}
  {2004})}\BibitemShut {NoStop}%
\bibitem [{\citenamefont {Goncalves}\ \emph
  {et~al.}(2001{\natexlab{a}})\citenamefont {Goncalves}, \citenamefont
  {Megretski},\ and\ \citenamefont {Dahleh}}]{Goncalves2001}%
  \BibitemOpen
  \bibfield  {author} {\bibinfo {author} {\bibfnamefont {J.~M.}\ \bibnamefont
  {Goncalves}}, \bibinfo {author} {\bibfnamefont {A.}~\bibnamefont
  {Megretski}},\ and\ \bibinfo {author} {\bibfnamefont {M.~A.}\ \bibnamefont
  {Dahleh}},\ }\bibfield  {title} {\bibinfo {title} {Global stability of relay
  feedback systems},\ }\href@noop {} {\bibfield  {journal} {\bibinfo  {journal}
  {IEEE Transactions on Automatic Control}\ }\textbf {\bibinfo {volume} {46}},\
  \bibinfo {pages} {550} (\bibinfo {year} {2001}{\natexlab{a}})}\BibitemShut
  {NoStop}%
\bibitem [{\citenamefont {Zhanybai}\ \emph {et~al.}(2015)\citenamefont
  {Zhanybai}, \citenamefont {Erik}, \citenamefont {Vasily},\ and\ \citenamefont
  {Roman}}]{ZHUSUBALIYEV2015}%
  \BibitemOpen
  \bibfield  {author} {\bibinfo {author} {\bibfnamefont {T.~Z.}\ \bibnamefont
  {Zhanybai}}, \bibinfo {author} {\bibfnamefont {M.}~\bibnamefont {Erik}},
  \bibinfo {author} {\bibfnamefont {G.~R.}\ \bibnamefont {Vasily}},\ and\
  \bibinfo {author} {\bibfnamefont {A.~N.}\ \bibnamefont {Roman}},\ }\bibfield
  {title} {\bibinfo {title} {Multistability and hidden attractors in a relay
  system with hysteresis},\ }\href@noop {} {\bibfield  {journal} {\bibinfo
  {journal} {Physica D: Nonlinear Phenomena}\ }\textbf {\bibinfo {volume}
  {306}},\ \bibinfo {pages} {6} (\bibinfo {year} {2015})}\BibitemShut {NoStop}%
\bibitem [{\citenamefont {Sieber}(2006)}]{Sieber2006}%
  \BibitemOpen
  \bibfield  {author} {\bibinfo {author} {\bibfnamefont {J.}~\bibnamefont
  {Sieber}},\ }\bibfield  {title} {\bibinfo {title} {Dynamics of delayed relay
  systems},\ }\href@noop {} {\bibfield  {journal} {\bibinfo  {journal}
  {Nonlinearity}\ }\textbf {\bibinfo {volume} {19}},\ \bibinfo {pages} {2489}
  (\bibinfo {year} {2006})}\BibitemShut {NoStop}%
\bibitem [{\citenamefont {Kalm{\'a}r-Nagy}\ \emph {et~al.}(2011)\citenamefont
  {Kalm{\'a}r-Nagy}, \citenamefont {Wahi},\ and\ \citenamefont
  {Halder}}]{Nagy2011}%
  \BibitemOpen
  \bibfield  {author} {\bibinfo {author} {\bibfnamefont {T.}~\bibnamefont
  {Kalm{\'a}r-Nagy}}, \bibinfo {author} {\bibfnamefont {P.}~\bibnamefont
  {Wahi}},\ and\ \bibinfo {author} {\bibfnamefont {A.}~\bibnamefont {Halder}},\
  }\bibfield  {title} {\bibinfo {title} {Dynamics of a hysteretic relay
  oscillator with periodic forcing},\ }\href@noop {} {\bibfield  {journal}
  {\bibinfo  {journal} {SIAM Journal on Applied Dynamical Systems}\ }\textbf
  {\bibinfo {volume} {10}},\ \bibinfo {pages} {403} (\bibinfo {year}
  {2011})}\BibitemShut {NoStop}%
\bibitem [{\citenamefont {Lelkes}\ and\ \citenamefont
  {Kalm{\'a}r-Nagy}(2021)}]{lelkes2021}%
  \BibitemOpen
  \bibfield  {author} {\bibinfo {author} {\bibfnamefont {J.}~\bibnamefont
  {Lelkes}}\ and\ \bibinfo {author} {\bibfnamefont {T.}~\bibnamefont
  {Kalm{\'a}r-Nagy}},\ }\bibfield  {title} {\bibinfo {title} {Analysis of a
  mass-spring-relay system with periodic forcing},\ }\href@noop {} {\bibfield
  {journal} {\bibinfo  {journal} {Nonlinear Dynamics}\ } (\bibinfo {year}
  {2021})}\BibitemShut {NoStop}%
\bibitem [{\citenamefont {{P}reisach}(1935)}]{Preisach1935}%
  \BibitemOpen
  \bibfield  {author} {\bibinfo {author} {\bibfnamefont {F.}~\bibnamefont
  {{P}reisach}},\ }\bibfield  {title} {\bibinfo {title} {{Ü}ber die
  magnetische {N}achwirkung},\ }\href@noop {} {\bibfield  {journal} {\bibinfo
  {journal} {Zeitschrift für Physik}\ }\textbf {\bibinfo {volume} {94}},\
  \bibinfo {pages} {277} (\bibinfo {year} {1935})}\BibitemShut {NoStop}%
\bibitem [{\citenamefont {Mayergoyz}(2003)}]{Mayergoyz2003}%
  \BibitemOpen
  \bibfield  {author} {\bibinfo {author} {\bibfnamefont {I.~D.}\ \bibnamefont
  {Mayergoyz}},\ }\href@noop {} {\emph {\bibinfo {title} {Mathematical models
  of hysteresis and their applications}}}\ (\bibinfo  {publisher} {Academic
  Press},\ \bibinfo {address} {New York},\ \bibinfo {year} {2003})\BibitemShut
  {NoStop}%
\bibitem [{\citenamefont {Ruderman}\ \emph {et~al.}(2009)\citenamefont
  {Ruderman}, \citenamefont {Hoffmann},\ and\ \citenamefont
  {Bertram}}]{Ruderman2009}%
  \BibitemOpen
  \bibfield  {author} {\bibinfo {author} {\bibfnamefont {M.}~\bibnamefont
  {Ruderman}}, \bibinfo {author} {\bibfnamefont {F.}~\bibnamefont {Hoffmann}},\
  and\ \bibinfo {author} {\bibfnamefont {T.}~\bibnamefont {Bertram}},\
  }\bibfield  {title} {\bibinfo {title} {Modeling and identification of elastic
  robot joints with hysteresis and backlash},\ }\href@noop {} {\bibfield
  {journal} {\bibinfo  {journal} {IEEE Transactions on Industrial Electronics}\
  }\textbf {\bibinfo {volume} {56}},\ \bibinfo {pages} {3840} (\bibinfo {year}
  {2009})}\BibitemShut {NoStop}%
\bibitem [{\citenamefont {Lamba}\ \emph {et~al.}(1997)\citenamefont {Lamba},
  \citenamefont {Grinfeld}, \citenamefont {McKee},\ and\ \citenamefont
  {Simpson}}]{Lamba1997}%
  \BibitemOpen
  \bibfield  {author} {\bibinfo {author} {\bibfnamefont {H.}~\bibnamefont
  {Lamba}}, \bibinfo {author} {\bibfnamefont {M.}~\bibnamefont {Grinfeld}},
  \bibinfo {author} {\bibfnamefont {S.}~\bibnamefont {McKee}},\ and\ \bibinfo
  {author} {\bibfnamefont {R.}~\bibnamefont {Simpson}},\ }\bibfield  {title}
  {\bibinfo {title} {Subharmonic ferroresonance in an lcr circuit with
  hysteresis},\ }\href@noop {} {\bibfield  {journal} {\bibinfo  {journal} {IEEE
  Trans. Magn.}\ }\textbf {\bibinfo {volume} {33}},\ \bibinfo {pages} {2495}
  (\bibinfo {year} {1997})}\BibitemShut {NoStop}%
\bibitem [{\citenamefont {Rezaei-Zare}\ \emph {et~al.}(2007)\citenamefont
  {Rezaei-Zare}, \citenamefont {Sanaye-Pasand}, \citenamefont {Mohseni},
  \citenamefont {Farhangi},\ and\ \citenamefont {Iravani}}]{Rezaei-Zare2007}%
  \BibitemOpen
  \bibfield  {author} {\bibinfo {author} {\bibfnamefont {A.}~\bibnamefont
  {Rezaei-Zare}}, \bibinfo {author} {\bibfnamefont {M.}~\bibnamefont
  {Sanaye-Pasand}}, \bibinfo {author} {\bibfnamefont {H.}~\bibnamefont
  {Mohseni}}, \bibinfo {author} {\bibfnamefont {S.}~\bibnamefont {Farhangi}},\
  and\ \bibinfo {author} {\bibfnamefont {R.}~\bibnamefont {Iravani}},\
  }\bibfield  {title} {\bibinfo {title} {Analysis of ferroresonance modes in
  power transformers using {P}reisach-type hysteretic magnetizing inductance},\
  }\href@noop {} {\bibfield  {journal} {\bibinfo  {journal} {IEEE Trans. Power
  Delivery}\ }\textbf {\bibinfo {volume} {22}},\ \bibinfo {pages} {919}
  (\bibinfo {year} {2007})}\BibitemShut {NoStop}%
\bibitem [{\citenamefont {Donnagain}\ and\ \citenamefont
  {Rasskazov}(2006)}]{Donnagain2006}%
  \BibitemOpen
  \bibfield  {author} {\bibinfo {author} {\bibfnamefont {M.~O.}\ \bibnamefont
  {Donnagain}}\ and\ \bibinfo {author} {\bibfnamefont {O.}~\bibnamefont
  {Rasskazov}},\ }\bibfield  {title} {\bibinfo {title} {Numerical modelling of
  an iron pendulum in a magnetic field},\ }\href@noop {} {\bibfield  {journal}
  {\bibinfo  {journal} {Physica B}\ }\textbf {\bibinfo {volume} {372}},\
  \bibinfo {pages} {37} (\bibinfo {year} {2006})}\BibitemShut {NoStop}%
\bibitem [{\citenamefont {Radons}\ and\ \citenamefont
  {Zienert}(2013)}]{radons2013nonlinear}%
  \BibitemOpen
  \bibfield  {author} {\bibinfo {author} {\bibfnamefont {G.}~\bibnamefont
  {Radons}}\ and\ \bibinfo {author} {\bibfnamefont {A.}~\bibnamefont
  {Zienert}},\ }\bibfield  {title} {\bibinfo {title} {Nonlinear dynamics of
  complex hysteretic systems: Oscillator in a magnetic field},\ }\href@noop {}
  {\bibfield  {journal} {\bibinfo  {journal} {Eur. Phys. J. Special Topic}\
  }\textbf {\bibinfo {volume} {222}},\ \bibinfo {pages} {1675} (\bibinfo {year}
  {2013})}\BibitemShut {NoStop}%
\bibitem [{\citenamefont {{Kuhnen}}\ and\ \citenamefont
  {{Janocha}}(1999)}]{kuhnen1999}%
  \BibitemOpen
  \bibfield  {author} {\bibinfo {author} {\bibfnamefont {K.}~\bibnamefont
  {{Kuhnen}}}\ and\ \bibinfo {author} {\bibfnamefont {H.}~\bibnamefont
  {{Janocha}}},\ }\bibfield  {title} {\bibinfo {title} {Adaptive inverse
  control of piezoelectric actuators with hysteresis operators},\ }in\
  \href@noop {} {\emph {\bibinfo {booktitle} {1999 European Control Conference
  (ECC)}}}\ (\bibinfo {year} {1999})\ pp.\ \bibinfo {pages}
  {791--796}\BibitemShut {NoStop}%
\bibitem [{\citenamefont {Ruderman}\ \emph {et~al.}(2019)\citenamefont
  {Ruderman}, \citenamefont {Yamada},\ and\ \citenamefont
  {Fujimoto}}]{Ruderman2019}%
  \BibitemOpen
  \bibfield  {author} {\bibinfo {author} {\bibfnamefont {M.}~\bibnamefont
  {Ruderman}}, \bibinfo {author} {\bibfnamefont {S.}~\bibnamefont {Yamada}},\
  and\ \bibinfo {author} {\bibfnamefont {H.}~\bibnamefont {Fujimoto}},\
  }\bibfield  {title} {\bibinfo {title} {{Backlash Identification in Two-Mass
  Systems by Delayed Relay Feedback}},\ }\href@noop {} {\bibfield  {journal}
  {\bibinfo  {journal} {Journal of Dynamic Systems, Measurement, and Control}\
  }\textbf {\bibinfo {volume} {141}} (\bibinfo {year} {2019})}\BibitemShut
  {NoStop}%
\bibitem [{\citenamefont {Visintin}(2013)}]{visintin2013}%
  \BibitemOpen
  \bibfield  {author} {\bibinfo {author} {\bibfnamefont {A.}~\bibnamefont
  {Visintin}},\ }\href@noop {} {\emph {\bibinfo {title} {Differential Models of
  Hysteresis}}},\ Applied Mathematical Sciences\ (\bibinfo  {publisher}
  {Springer, Berlin},\ \bibinfo {year} {2013})\BibitemShut {NoStop}%
\bibitem [{\citenamefont {Krej\v{c}\'i}\ and\ \citenamefont
  {Kuhnen}(2001)}]{Krejci2001}%
  \BibitemOpen
  \bibfield  {author} {\bibinfo {author} {\bibfnamefont {P.}~\bibnamefont
  {Krej\v{c}\'i}}\ and\ \bibinfo {author} {\bibfnamefont {K.}~\bibnamefont
  {Kuhnen}},\ }\bibfield  {title} {\bibinfo {title} {Inverse control of systems
  with hysteresis and creep},\ }\href@noop {} {\bibfield  {journal} {\bibinfo
  {journal} {IEE Proc. Control Theory Appl.}\ }\textbf {\bibinfo {volume}
  {148}},\ \bibinfo {pages} {185} (\bibinfo {year} {2001})}\BibitemShut
  {NoStop}%
\bibitem [{\citenamefont {Riccardi}\ \emph {et~al.}(2014)\citenamefont
  {Riccardi}, \citenamefont {Naso}, \citenamefont {Turchiano},\ and\
  \citenamefont {Janocha}}]{Riccardi2014}%
  \BibitemOpen
  \bibfield  {author} {\bibinfo {author} {\bibfnamefont {L.}~\bibnamefont
  {Riccardi}}, \bibinfo {author} {\bibfnamefont {D.}~\bibnamefont {Naso}},
  \bibinfo {author} {\bibfnamefont {B.}~\bibnamefont {Turchiano}},\ and\
  \bibinfo {author} {\bibfnamefont {H.}~\bibnamefont {Janocha}},\ }\bibfield
  {title} {\bibinfo {title} {Design of linear feedback controllers for dynamic
  systems with hysteresis},\ }\href@noop {} {\bibfield  {journal} {\bibinfo
  {journal} {IEEE Transactions on Control Systems Technology}\ }\textbf
  {\bibinfo {volume} {22}},\ \bibinfo {pages} {1268} (\bibinfo {year}
  {2014})}\BibitemShut {NoStop}%
\bibitem [{\citenamefont {Imry}\ and\ \citenamefont {Ma}(1975)}]{Imry1975}%
  \BibitemOpen
  \bibfield  {author} {\bibinfo {author} {\bibfnamefont {Y.}~\bibnamefont
  {Imry}}\ and\ \bibinfo {author} {\bibfnamefont {S.}~\bibnamefont {Ma}},\
  }\bibfield  {title} {\bibinfo {title} {Random-field instability of the
  ordered state of continuous symmetry},\ }\href@noop {} {\bibfield  {journal}
  {\bibinfo  {journal} {Phys. Rev. Lett.}\ }\textbf {\bibinfo {volume} {35}},\
  \bibinfo {pages} {1399} (\bibinfo {year} {1975})}\BibitemShut {NoStop}%
\bibitem [{\citenamefont {Perkovic}\ \emph {et~al.}(1995)\citenamefont
  {Perkovic}, \citenamefont {Dahmen},\ and\ \citenamefont
  {Sethna}}]{Perkovic1995}%
  \BibitemOpen
  \bibfield  {author} {\bibinfo {author} {\bibfnamefont {O.}~\bibnamefont
  {Perkovic}}, \bibinfo {author} {\bibfnamefont {K.~A.}\ \bibnamefont
  {Dahmen}},\ and\ \bibinfo {author} {\bibfnamefont {J.~P.}\ \bibnamefont
  {Sethna}},\ }\bibfield  {title} {\bibinfo {title} {Avalanches, {B}arkhausen
  noise, and plain old criticality},\ }\href@noop {} {\bibfield  {journal}
  {\bibinfo  {journal} {Phys. Rev. Lett.}\ }\textbf {\bibinfo {volume} {75}},\
  \bibinfo {pages} {4528} (\bibinfo {year} {1995})}\BibitemShut {NoStop}%
\bibitem [{\citenamefont {Shukla}(1996)}]{Shukla1996}%
  \BibitemOpen
  \bibfield  {author} {\bibinfo {author} {\bibfnamefont {P.}~\bibnamefont
  {Shukla}},\ }\bibfield  {title} {\bibinfo {title} {Exact solution of
  zero-temperature hysteresis in a ferromagnetic {I}sing chain with quenched
  random fields},\ }\href@noop {} {\bibfield  {journal} {\bibinfo  {journal}
  {Physica A}\ }\textbf {\bibinfo {volume} {233}},\ \bibinfo {pages} {235}
  (\bibinfo {year} {1996})}\BibitemShut {NoStop}%
\bibitem [{\citenamefont {Sethna}\ \emph {et~al.}(2001)\citenamefont {Sethna},
  \citenamefont {Dahmen},\ and\ \citenamefont {Myers}}]{Sethna2001}%
  \BibitemOpen
  \bibfield  {author} {\bibinfo {author} {\bibfnamefont {J.~P.}\ \bibnamefont
  {Sethna}}, \bibinfo {author} {\bibfnamefont {K.~A.}\ \bibnamefont {Dahmen}},\
  and\ \bibinfo {author} {\bibfnamefont {C.~R.}\ \bibnamefont {Myers}},\
  }\bibfield  {title} {\bibinfo {title} {Crackling noise},\ }\href@noop {}
  {\bibfield  {journal} {\bibinfo  {journal} {Nature}\ }\textbf {\bibinfo
  {volume} {410}},\ \bibinfo {pages} {242} (\bibinfo {year}
  {2001})}\BibitemShut {NoStop}%
\bibitem [{\citenamefont {Zech}\ \emph {et~al.}(2020)\citenamefont {Zech},
  \citenamefont {Otto},\ and\ \citenamefont {Radons}}]{zech2020}%
  \BibitemOpen
  \bibfield  {author} {\bibinfo {author} {\bibfnamefont {P.}~\bibnamefont
  {Zech}}, \bibinfo {author} {\bibfnamefont {A.}~\bibnamefont {Otto}},\ and\
  \bibinfo {author} {\bibfnamefont {G.}~\bibnamefont {Radons}},\ }\bibfield
  {title} {\bibinfo {title} {Dynamics of a driven harmonic oscillator coupled
  to independent {I}sing spins in random fields},\ }\href@noop {} {\bibfield
  {journal} {\bibinfo  {journal} {Phys. Rev. E}\ }\textbf {\bibinfo {volume}
  {101}},\ \bibinfo {pages} {042217} (\bibinfo {year} {2020})}\BibitemShut
  {NoStop}%
\bibitem [{\citenamefont {Fowler}\ and\ \citenamefont
  {Puga}(1978)}]{fowler1978}%
  \BibitemOpen
  \bibfield  {author} {\bibinfo {author} {\bibfnamefont {M.}~\bibnamefont
  {Fowler}}\ and\ \bibinfo {author} {\bibfnamefont {M.~W.}\ \bibnamefont
  {Puga}},\ }\bibfield  {title} {\bibinfo {title} {Dimer gas model for
  tetracyanoquinodimethane (tcnq)},\ }\href@noop {} {\bibfield  {journal}
  {\bibinfo  {journal} {Phys. Rev. B}\ }\textbf {\bibinfo {volume} {18}},\
  \bibinfo {pages} {421} (\bibinfo {year} {1978})}\BibitemShut {NoStop}%
\bibitem [{\citenamefont {Ramos}\ \emph {et~al.}(2014)\citenamefont {Ramos},
  \citenamefont {Pichler}, \citenamefont {Daley},\ and\ \citenamefont
  {Zoller}}]{ramos2014}%
  \BibitemOpen
  \bibfield  {author} {\bibinfo {author} {\bibfnamefont {T.}~\bibnamefont
  {Ramos}}, \bibinfo {author} {\bibfnamefont {H.}~\bibnamefont {Pichler}},
  \bibinfo {author} {\bibfnamefont {A.~J.}\ \bibnamefont {Daley}},\ and\
  \bibinfo {author} {\bibfnamefont {P.}~\bibnamefont {Zoller}},\ }\bibfield
  {title} {\bibinfo {title} {Quantum spin dimers from chiral dissipation in
  cold-atom chains},\ }\href@noop {} {\bibfield  {journal} {\bibinfo  {journal}
  {Phys. Rev. Lett.}\ }\textbf {\bibinfo {volume} {113}},\ \bibinfo {pages}
  {237203} (\bibinfo {year} {2014})}\BibitemShut {NoStop}%
\bibitem [{\citenamefont {Visintin}(1994)}]{Visintin1994}%
  \BibitemOpen
  \bibfield  {author} {\bibinfo {author} {\bibfnamefont {A.}~\bibnamefont
  {Visintin}},\ }\href@noop {} {\emph {\bibinfo {title} {Differential models of
  hysteresis}}}\ (\bibinfo  {publisher} {Springer, Berlin},\ \bibinfo {year}
  {1994})\BibitemShut {NoStop}%
\bibitem [{\citenamefont {Krasnosel'skii}\ and\ \citenamefont
  {Pokrovskii}(1989)}]{krasnosel1989}%
  \BibitemOpen
  \bibfield  {author} {\bibinfo {author} {\bibfnamefont {M.~A.}\ \bibnamefont
  {Krasnosel'skii}}\ and\ \bibinfo {author} {\bibfnamefont {A.~V.}\
  \bibnamefont {Pokrovskii}},\ }\href@noop {} {\emph {\bibinfo {title} {Systems
  with Hysteresis}}},\ Springer Series in Solid-State Sciences\ (\bibinfo
  {publisher} {Springer, New York},\ \bibinfo {year} {1989})\BibitemShut
  {NoStop}%
\bibitem [{\citenamefont {Hassani}\ \emph {et~al.}(2014)\citenamefont
  {Hassani}, \citenamefont {Tjahjowidodo},\ and\ \citenamefont
  {Do}}]{HASSANI2014}%
  \BibitemOpen
  \bibfield  {author} {\bibinfo {author} {\bibfnamefont {V.}~\bibnamefont
  {Hassani}}, \bibinfo {author} {\bibfnamefont {T.}~\bibnamefont
  {Tjahjowidodo}},\ and\ \bibinfo {author} {\bibfnamefont {T.~N.}\ \bibnamefont
  {Do}},\ }\bibfield  {title} {\bibinfo {title} {A survey on hysteresis
  modeling, identification and control},\ }\href@noop {} {\bibfield  {journal}
  {\bibinfo  {journal} {Mechanical Systems and Signal Processing}\ }\textbf
  {\bibinfo {volume} {49}},\ \bibinfo {pages} {209} (\bibinfo {year}
  {2014})}\BibitemShut {NoStop}%
\bibitem [{\citenamefont {Al~Janaideh}\ \emph {et~al.}(2008)\citenamefont
  {Al~Janaideh}, \citenamefont {Mao}, \citenamefont {Rakheja}, \citenamefont
  {Xie},\ and\ \citenamefont {Su}}]{aljanaideh2008}%
  \BibitemOpen
  \bibfield  {author} {\bibinfo {author} {\bibfnamefont {M.}~\bibnamefont
  {Al~Janaideh}}, \bibinfo {author} {\bibfnamefont {J.}~\bibnamefont {Mao}},
  \bibinfo {author} {\bibfnamefont {S.}~\bibnamefont {Rakheja}}, \bibinfo
  {author} {\bibfnamefont {W.}~\bibnamefont {Xie}},\ and\ \bibinfo {author}
  {\bibfnamefont {C.}~\bibnamefont {Su}},\ }\bibfield  {title} {\bibinfo
  {title} {Generalized {P}randtl-{I}shlinskii hysteresis model: Hysteresis
  modeling and its inverse for compensation in smart actuators},\ }in\
  \href@noop {} {\emph {\bibinfo {booktitle} {2008 47th IEEE Conference on
  Decision and Control}}}\ (\bibinfo {year} {2008})\ pp.\ \bibinfo {pages}
  {5182--5187}\BibitemShut {NoStop}%
\bibitem [{\citenamefont {Vives}\ \emph {et~al.}(2005)\citenamefont {Vives},
  \citenamefont {Rosinberg},\ and\ \citenamefont {Tarjus}}]{vives2005}%
  \BibitemOpen
  \bibfield  {author} {\bibinfo {author} {\bibfnamefont {E.}~\bibnamefont
  {Vives}}, \bibinfo {author} {\bibfnamefont {M.~L.}\ \bibnamefont
  {Rosinberg}},\ and\ \bibinfo {author} {\bibfnamefont {G.}~\bibnamefont
  {Tarjus}},\ }\bibfield  {title} {\bibinfo {title} {Hysteresis and avalanches
  in the $t=0$ random-field {I}sing model with two-spin-flip dynamics},\
  }\href@noop {} {\bibfield  {journal} {\bibinfo  {journal} {Phys. Rev. B}\
  }\textbf {\bibinfo {volume} {71}},\ \bibinfo {pages} {134424} (\bibinfo
  {year} {2005})}\BibitemShut {NoStop}%
\bibitem [{\citenamefont {Salvat-Pujol}\ \emph {et~al.}(2009)\citenamefont
  {Salvat-Pujol}, \citenamefont {Vives},\ and\ \citenamefont
  {Rosinberg}}]{salvat2009}%
  \BibitemOpen
  \bibfield  {author} {\bibinfo {author} {\bibfnamefont {F.}~\bibnamefont
  {Salvat-Pujol}}, \bibinfo {author} {\bibfnamefont {E.}~\bibnamefont
  {Vives}},\ and\ \bibinfo {author} {\bibfnamefont {M.~L.}\ \bibnamefont
  {Rosinberg}},\ }\bibfield  {title} {\bibinfo {title} {Hysteresis in the $t=0$
  random-field {I}sing model: Beyond metastable dynamics},\ }\href@noop {}
  {\bibfield  {journal} {\bibinfo  {journal} {Phys. Rev. E}\ }\textbf {\bibinfo
  {volume} {79}},\ \bibinfo {pages} {061116} (\bibinfo {year}
  {2009})}\BibitemShut {NoStop}%
\bibitem [{\citenamefont {Dahmen}\ and\ \citenamefont
  {Sethna}(1996)}]{Dahmen1996}%
  \BibitemOpen
  \bibfield  {author} {\bibinfo {author} {\bibfnamefont {K.}~\bibnamefont
  {Dahmen}}\ and\ \bibinfo {author} {\bibfnamefont {J.~P.}\ \bibnamefont
  {Sethna}},\ }\bibfield  {title} {\bibinfo {title} {Hysteresis, avalanches,
  and disorder-induced critical scaling: A renormalization-group approach},\
  }\href@noop {} {\bibfield  {journal} {\bibinfo  {journal} {Phys. Rev. B}\
  }\textbf {\bibinfo {volume} {53}},\ \bibinfo {pages} {14872} (\bibinfo {year}
  {1996})}\BibitemShut {NoStop}%
\bibitem [{\citenamefont {di~Bernardo}\ \emph {et~al.}(2008)\citenamefont
  {di~Bernardo}, \citenamefont {Budd}, \citenamefont {Champneys},\ and\
  \citenamefont {Kowalczyk}}]{bernardo2008}%
  \BibitemOpen
  \bibfield  {author} {\bibinfo {author} {\bibfnamefont {M.}~\bibnamefont
  {di~Bernardo}}, \bibinfo {author} {\bibfnamefont {C.}~\bibnamefont {Budd}},
  \bibinfo {author} {\bibfnamefont {A.~R.}\ \bibnamefont {Champneys}},\ and\
  \bibinfo {author} {\bibfnamefont {P.}~\bibnamefont {Kowalczyk}},\ }\href@noop
  {} {\emph {\bibinfo {title} {Piecewise-smooth Dynamical Systems: Theory and
  Applications}}},\ \bibinfo {series} {Applied Mathematical Sciences}, Vol.\
  \bibinfo {volume} {163}\ (\bibinfo  {publisher} {Springer, London},\ \bibinfo
  {year} {2008})\BibitemShut {NoStop}%
\bibitem [{\citenamefont {Goncalves}\ \emph
  {et~al.}(2001{\natexlab{b}})\citenamefont {Goncalves}, \citenamefont
  {Megretski},\ and\ \citenamefont {Dahleh}}]{Goncalvez2001}%
  \BibitemOpen
  \bibfield  {author} {\bibinfo {author} {\bibfnamefont {J.~M.}\ \bibnamefont
  {Goncalves}}, \bibinfo {author} {\bibfnamefont {A.}~\bibnamefont
  {Megretski}},\ and\ \bibinfo {author} {\bibfnamefont {M.~A.}\ \bibnamefont
  {Dahleh}},\ }\bibfield  {title} {\bibinfo {title} {Global stability of relay
  feedback systems},\ }\href@noop {} {\bibfield  {journal} {\bibinfo  {journal}
  {IEEE Transactions on Automatic Control}\ }\textbf {\bibinfo {volume} {46}},\
  \bibinfo {pages} {550} (\bibinfo {year} {2001}{\natexlab{b}})}\BibitemShut
  {NoStop}%
\bibitem [{\citenamefont {Theodossiades}\ and\ \citenamefont
  {Natsiavas}(2000)}]{theodossiades2000}%
  \BibitemOpen
  \bibfield  {author} {\bibinfo {author} {\bibfnamefont {S.}~\bibnamefont
  {Theodossiades}}\ and\ \bibinfo {author} {\bibfnamefont {S.}~\bibnamefont
  {Natsiavas}},\ }\bibfield  {title} {\bibinfo {title} {Non-linear dynamics of
  gear-pair systems with periodic stiffness and backlash},\ }\href@noop {}
  {\bibfield  {journal} {\bibinfo  {journal} {Journal of Sound and Vibration}\
  }\textbf {\bibinfo {volume} {229}},\ \bibinfo {pages} {287} (\bibinfo {year}
  {2000})}\BibitemShut {NoStop}%
\bibitem [{\citenamefont {Galvanetto}(2001)}]{galvanetto2001}%
  \BibitemOpen
  \bibfield  {author} {\bibinfo {author} {\bibfnamefont {U.}~\bibnamefont
  {Galvanetto}},\ }\bibfield  {title} {\bibinfo {title} {Some discontinuous
  bifurcations in a two-blocks stick-slip system},\ }\href@noop {} {\bibfield
  {journal} {\bibinfo  {journal} {Journal of Sound and Vibration}\ }\textbf
  {\bibinfo {volume} {248}},\ \bibinfo {pages} {653} (\bibinfo {year}
  {2001})}\BibitemShut {NoStop}%
\bibitem [{\citenamefont {Zhao}\ and\ \citenamefont
  {Dankowicz}(2006)}]{Zhao2005}%
  \BibitemOpen
  \bibfield  {author} {\bibinfo {author} {\bibfnamefont {X.}~\bibnamefont
  {Zhao}}\ and\ \bibinfo {author} {\bibfnamefont {H.}~\bibnamefont
  {Dankowicz}},\ }\bibfield  {title} {\bibinfo {title} {Characterization of
  intermittent contact in tapping mode atomic force microscopy},\ }\href@noop
  {} {\bibfield  {journal} {\bibinfo  {journal} {Journal of Computational and
  Nonlinear Dynamics}\ }\textbf {\bibinfo {volume} {1}},\ \bibinfo {pages}
  {109} (\bibinfo {year} {2006})}\BibitemShut {NoStop}%
\bibitem [{\citenamefont {Mayergoyz}(1991)}]{Mayergoyz1991a}%
  \BibitemOpen
  \bibfield  {author} {\bibinfo {author} {\bibfnamefont {I.~D.}\ \bibnamefont
  {Mayergoyz}},\ }\href@noop {} {\emph {\bibinfo {title} {Mathematical models
  of hysteresis}}}\ (\bibinfo  {publisher} {Springer, New York},\ \bibinfo
  {year} {1991})\BibitemShut {NoStop}%
\bibitem [{\citenamefont {Bennetin}\ \emph {et~al.}(1980)\citenamefont
  {Bennetin}, \citenamefont {Galgani}, \citenamefont {Giorgilli},\ and\
  \citenamefont {Strelcyn}}]{1980bennetin}%
  \BibitemOpen
  \bibfield  {author} {\bibinfo {author} {\bibfnamefont {G.}~\bibnamefont
  {Bennetin}}, \bibinfo {author} {\bibfnamefont {L.}~\bibnamefont {Galgani}},
  \bibinfo {author} {\bibfnamefont {A.}~\bibnamefont {Giorgilli}},\ and\
  \bibinfo {author} {\bibfnamefont {J.}~\bibnamefont {Strelcyn}},\ }\bibfield
  {title} {\bibinfo {title} {{{L}yapunov characteristic exponents for smooth
  dynamical systems and for Hamiltonian systems; a method for computing all of
  them. Part 1: Theory. Part 2: Numerical applications}},\ }\href@noop {}
  {\bibfield  {journal} {\bibinfo  {journal} {Meccanica}\ }\textbf {\bibinfo
  {volume} {15}},\ \bibinfo {pages} {9} (\bibinfo {year} {1980})}\BibitemShut
  {NoStop}%
\bibitem [{\citenamefont {Nordmark}(1991)}]{nordmark1991}%
  \BibitemOpen
  \bibfield  {author} {\bibinfo {author} {\bibfnamefont {A.~B.}\ \bibnamefont
  {Nordmark}},\ }\bibfield  {title} {\bibinfo {title} {Non-periodic motion
  caused by grazing incidence in an impact oscillator},\ }\href@noop {}
  {\bibfield  {journal} {\bibinfo  {journal} {Journal of Sound and Vibration}\
  }\textbf {\bibinfo {volume} {145}},\ \bibinfo {pages} {279} (\bibinfo {year}
  {1991})}\BibitemShut {NoStop}%
\bibitem [{\citenamefont {M\"uller}(1995)}]{mueller1995}%
  \BibitemOpen
  \bibfield  {author} {\bibinfo {author} {\bibfnamefont {P.~C.}\ \bibnamefont
  {M\"uller}},\ }\bibfield  {title} {\bibinfo {title} {Calculation of
  {L}yapunov exponents for dynamic systems withdiscontinuities},\ }\href@noop
  {} {\bibfield  {journal} {\bibinfo  {journal} {Chaos, Solitons \& Fractals}\
  }\textbf {\bibinfo {volume} {5}},\ \bibinfo {pages} {1671} (\bibinfo {year}
  {1995})}\BibitemShut {NoStop}%
\bibitem [{\citenamefont {Dankowicz}\ and\ \citenamefont
  {Nordmark}(2000)}]{dankowicz2000}%
  \BibitemOpen
  \bibfield  {author} {\bibinfo {author} {\bibfnamefont {H.}~\bibnamefont
  {Dankowicz}}\ and\ \bibinfo {author} {\bibfnamefont {A.~B.}\ \bibnamefont
  {Nordmark}},\ }\bibfield  {title} {\bibinfo {title} {On the origin and
  bifurcations of stick-slip oscillations},\ }\href@noop {} {\bibfield
  {journal} {\bibinfo  {journal} {Physica D: Nonlinear Phenomena}\ }\textbf
  {\bibinfo {volume} {136}},\ \bibinfo {pages} {280} (\bibinfo {year}
  {2000})}\BibitemShut {NoStop}%
\bibitem [{\citenamefont {di~Bernardo}\ \emph {et~al.}(2001)\citenamefont
  {di~Bernardo}, \citenamefont {Budd},\ and\ \citenamefont
  {Champneys}}]{bernardo20011}%
  \BibitemOpen
  \bibfield  {author} {\bibinfo {author} {\bibfnamefont {M.}~\bibnamefont
  {di~Bernardo}}, \bibinfo {author} {\bibfnamefont {C.~J.}\ \bibnamefont
  {Budd}},\ and\ \bibinfo {author} {\bibfnamefont {A.~R.}\ \bibnamefont
  {Champneys}},\ }\bibfield  {title} {\bibinfo {title} {Grazing and
  border-collision in piecewise-smooth systems: A unified analytical
  framework},\ }\href@noop {} {\bibfield  {journal} {\bibinfo  {journal} {Phys.
  Rev. Lett.}\ }\textbf {\bibinfo {volume} {86}},\ \bibinfo {pages} {2553}
  (\bibinfo {year} {2001})}\BibitemShut {NoStop}%
\bibitem [{\citenamefont {Geist}\ \emph {et~al.}(1990)\citenamefont {Geist},
  \citenamefont {Parlitz},\ and\ \citenamefont {Lauterborn}}]{geist1990}%
  \BibitemOpen
  \bibfield  {author} {\bibinfo {author} {\bibfnamefont {K.}~\bibnamefont
  {Geist}}, \bibinfo {author} {\bibfnamefont {U.}~\bibnamefont {Parlitz}},\
  and\ \bibinfo {author} {\bibfnamefont {W.}~\bibnamefont {Lauterborn}},\
  }\bibfield  {title} {\bibinfo {title} {Comparison of different methods for
  computing {L}yapunov exponents},\ }\href@noop {} {\bibfield  {journal}
  {\bibinfo  {journal} {Progress of Theoretical Physics}\ }\textbf {\bibinfo
  {volume} {83}},\ \bibinfo {pages} {875} (\bibinfo {year} {1990})}\BibitemShut
  {NoStop}%
\bibitem [{\citenamefont {Dieci}\ \emph {et~al.}(1997)\citenamefont {Dieci},
  \citenamefont {Russell},\ and\ \citenamefont {Van~Vleck}}]{dieci1997}%
  \BibitemOpen
  \bibfield  {author} {\bibinfo {author} {\bibfnamefont {L.}~\bibnamefont
  {Dieci}}, \bibinfo {author} {\bibfnamefont {R.~D.}\ \bibnamefont {Russell}},\
  and\ \bibinfo {author} {\bibfnamefont {E.~S.}\ \bibnamefont {Van~Vleck}},\
  }\bibfield  {title} {\bibinfo {title} {On the compuation of {L}yapunov
  exponents for continuous dynamical systems},\ }\href@noop {} {\bibfield
  {journal} {\bibinfo  {journal} {SIAM Journal on Numerical Analysis}\ }\textbf
  {\bibinfo {volume} {34}},\ \bibinfo {pages} {402} (\bibinfo {year}
  {1997})}\BibitemShut {NoStop}%
\bibitem [{\citenamefont {Brokate}\ and\ \citenamefont
  {Sprekels}(1996)}]{Brokate1996}%
  \BibitemOpen
  \bibfield  {author} {\bibinfo {author} {\bibfnamefont {M.}~\bibnamefont
  {Brokate}}\ and\ \bibinfo {author} {\bibfnamefont {J.}~\bibnamefont
  {Sprekels}},\ }\href@noop {} {\emph {\bibinfo {title} {Hysteresis and phase
  transitions}}}\ (\bibinfo  {publisher} {Springer, New York},\ \bibinfo {year}
  {1996})\BibitemShut {NoStop}%
\bibitem [{\citenamefont {Dimian}\ and\ \citenamefont
  {Andrei}(2013)}]{dimian2013}%
  \BibitemOpen
  \bibfield  {author} {\bibinfo {author} {\bibfnamefont {M.}~\bibnamefont
  {Dimian}}\ and\ \bibinfo {author} {\bibfnamefont {P.}~\bibnamefont
  {Andrei}},\ }\href@noop {} {\emph {\bibinfo {title} {Noise-Driven Phenomena
  in Hysteretic Systems}}},\ Signals and Communication Technology\ (\bibinfo
  {publisher} {Springer, New York},\ \bibinfo {year} {2013})\BibitemShut
  {NoStop}%
\bibitem [{Note1()}]{Note1}%
  \BibitemOpen
  \bibinfo {note} {In fact, if $ q(t) $ is an extremum (the finger touches one
  of the walls with velocity zero) this corresponds to a grazing intersection,
  which needs indeed to be investigate in more detail.}\BibitemShut {Stop}%
\bibitem [{\citenamefont {Radons}\ \emph {et~al.}(2009)\citenamefont {Radons},
  \citenamefont {Yang}, \citenamefont {Wang},\ and\ \citenamefont
  {Fu}}]{radons2009}%
  \BibitemOpen
  \bibfield  {author} {\bibinfo {author} {\bibfnamefont {G.}~\bibnamefont
  {Radons}}, \bibinfo {author} {\bibfnamefont {H.-L.}\ \bibnamefont {Yang}},
  \bibinfo {author} {\bibfnamefont {J.}~\bibnamefont {Wang}},\ and\ \bibinfo
  {author} {\bibfnamefont {J.-F.}\ \bibnamefont {Fu}},\ }\bibfield  {title}
  {\bibinfo {title} {Complex behavior of simple maps with fluctuating delay
  times},\ }\href@noop {} {\bibfield  {journal} {\bibinfo  {journal} {The
  European Physical Journal B}\ }\textbf {\bibinfo {volume} {71}},\ \bibinfo
  {pages} {111} (\bibinfo {year} {2009})}\BibitemShut {NoStop}%
\bibitem [{\citenamefont {Otto}\ and\ \citenamefont {Radons}(2010)}]{otto2010}%
  \BibitemOpen
  \bibfield  {author} {\bibinfo {author} {\bibfnamefont {A.}~\bibnamefont
  {Otto}}\ and\ \bibinfo {author} {\bibfnamefont {G.}~\bibnamefont {Radons}},\
  }\bibfield  {title} {\bibinfo {title} {{L}yapunov spectrum of linear delay
  differential equations with time-varying delay},\ }\href@noop {} {\bibfield
  {journal} {\bibinfo  {journal} {IFAC Proceedings Volumes}\ }\textbf {\bibinfo
  {volume} {43}},\ \bibinfo {pages} {206} (\bibinfo {year} {2010})}\BibitemShut
  {NoStop}%
\bibitem [{\citenamefont {{Wang}}\ \emph {et~al.}(2011)\citenamefont {{Wang}},
  \citenamefont {{Radons}},\ and\ \citenamefont {{Yang}}}]{wang2011}%
  \BibitemOpen
  \bibfield  {author} {\bibinfo {author} {\bibfnamefont {J.}~\bibnamefont
  {{Wang}}}, \bibinfo {author} {\bibfnamefont {G.}~\bibnamefont {{Radons}}},\
  and\ \bibinfo {author} {\bibfnamefont {H.-L.}\ \bibnamefont {{Yang}}},\
  }\href@noop {} {\bibinfo {title} {Dimensional collapse and fractal attractors
  of a system with fluctuating delay times}} (\bibinfo {year} {2011}),\ \Eprint
  {https://arxiv.org/abs/1112.1269} {arXiv:1112.1269 [nlin.CD]} \BibitemShut
  {NoStop}%
\bibitem [{\citenamefont {Grassberger}\ and\ \citenamefont
  {Procaccia}(1983)}]{grassberger1983}%
  \BibitemOpen
  \bibfield  {author} {\bibinfo {author} {\bibfnamefont {P.}~\bibnamefont
  {Grassberger}}\ and\ \bibinfo {author} {\bibfnamefont {I.}~\bibnamefont
  {Procaccia}},\ }\bibfield  {title} {\bibinfo {title} {Measuring the
  strangeness of strange attractors},\ }\href@noop {} {\bibfield  {journal}
  {\bibinfo  {journal} {Physica D: Nonlinear Phenomena}\ }\textbf {\bibinfo
  {volume} {9}},\ \bibinfo {pages} {189} (\bibinfo {year} {1983})}\BibitemShut
  {NoStop}%
\bibitem [{\citenamefont {Kaplan}\ and\ \citenamefont
  {Yorke}(1979)}]{kaplan1979}%
  \BibitemOpen
  \bibfield  {author} {\bibinfo {author} {\bibfnamefont {J.~L.}\ \bibnamefont
  {Kaplan}}\ and\ \bibinfo {author} {\bibfnamefont {J.~A.}\ \bibnamefont
  {Yorke}},\ }\bibfield  {title} {\bibinfo {title} {Chaotic behavior of
  multidimensional difference equations},\ }in\ \href@noop {} {\emph {\bibinfo
  {booktitle} {Functional differential equations and approximation of fixed
  points}}}\ (\bibinfo  {publisher} {Springer},\ \bibinfo {year} {1979})\ pp.\
  \bibinfo {pages} {204--227}\BibitemShut {NoStop}%
\bibitem [{\citenamefont {Farmer}\ \emph {et~al.}(1983)\citenamefont {Farmer},
  \citenamefont {Ott},\ and\ \citenamefont {Yorke}}]{farmer1983}%
  \BibitemOpen
  \bibfield  {author} {\bibinfo {author} {\bibfnamefont {J.~D.}\ \bibnamefont
  {Farmer}}, \bibinfo {author} {\bibfnamefont {E.}~\bibnamefont {Ott}},\ and\
  \bibinfo {author} {\bibfnamefont {J.~A.}\ \bibnamefont {Yorke}},\ }\bibfield
  {title} {\bibinfo {title} {The dimension of chaotic attractors},\ }\href@noop
  {} {\bibfield  {journal} {\bibinfo  {journal} {Physica D: Nonlinear
  Phenomena}\ }\textbf {\bibinfo {volume} {7}},\ \bibinfo {pages} {153}
  (\bibinfo {year} {1983})}\BibitemShut {NoStop}%
\bibitem [{\citenamefont {Falconer}(2013)}]{falconer2013}%
  \BibitemOpen
  \bibfield  {author} {\bibinfo {author} {\bibfnamefont {K.}~\bibnamefont
  {Falconer}},\ }\href@noop {} {\emph {\bibinfo {title} {Fractal Geometry:
  Mathematical Foundations and Applications}}}\ (\bibinfo  {publisher}
  {Wiley},\ \bibinfo {year} {2013})\BibitemShut {NoStop}%
\bibitem [{\citenamefont {Budd}\ and\ \citenamefont {Dux}(1994)}]{budd1994}%
  \BibitemOpen
  \bibfield  {author} {\bibinfo {author} {\bibfnamefont {C.}~\bibnamefont
  {Budd}}\ and\ \bibinfo {author} {\bibfnamefont {F.}~\bibnamefont {Dux}},\
  }\bibfield  {title} {\bibinfo {title} {Intermittency in impact oscillators
  close to resonance},\ }\href@noop {} {\bibfield  {journal} {\bibinfo
  {journal} {Nonlinearity}\ }\textbf {\bibinfo {volume} {7}},\ \bibinfo {pages}
  {1191} (\bibinfo {year} {1994})}\BibitemShut {NoStop}%
\bibitem [{\citenamefont {Chin}\ \emph {et~al.}(1994)\citenamefont {Chin},
  \citenamefont {Ott}, \citenamefont {Nusse},\ and\ \citenamefont
  {Grebogi}}]{chin1994}%
  \BibitemOpen
  \bibfield  {author} {\bibinfo {author} {\bibfnamefont {W.}~\bibnamefont
  {Chin}}, \bibinfo {author} {\bibfnamefont {E.}~\bibnamefont {Ott}}, \bibinfo
  {author} {\bibfnamefont {H.~E.}\ \bibnamefont {Nusse}},\ and\ \bibinfo
  {author} {\bibfnamefont {C.}~\bibnamefont {Grebogi}},\ }\bibfield  {title}
  {\bibinfo {title} {Grazing bifurcations in impact oscillators},\ }\href@noop
  {} {\bibfield  {journal} {\bibinfo  {journal} {Phys. Rev. E}\ }\textbf
  {\bibinfo {volume} {50}},\ \bibinfo {pages} {4427} (\bibinfo {year}
  {1994})}\BibitemShut {NoStop}%
\bibitem [{\citenamefont {Foale}\ and\ \citenamefont
  {Bishop}(1994)}]{foale1994}%
  \BibitemOpen
  \bibfield  {author} {\bibinfo {author} {\bibfnamefont {S.}~\bibnamefont
  {Foale}}\ and\ \bibinfo {author} {\bibfnamefont {S.~R.}\ \bibnamefont
  {Bishop}},\ }\bibfield  {title} {\bibinfo {title} {Bifurcations in impact
  oscillations},\ }\href@noop {} {\bibfield  {journal} {\bibinfo  {journal}
  {Nonlinear Dynamics}\ }\textbf {\bibinfo {volume} {6}},\ \bibinfo {pages}
  {285} (\bibinfo {year} {1994})}\BibitemShut {NoStop}%
\bibitem [{\citenamefont {Radons}\ \emph {et~al.}(2008)\citenamefont {Radons},
  \citenamefont {He\ss{}e}, \citenamefont {Lange},\ and\ \citenamefont
  {Schubert}}]{radons2008dynamics}%
  \BibitemOpen
  \bibfield  {author} {\bibinfo {author} {\bibfnamefont {G.}~\bibnamefont
  {Radons}}, \bibinfo {author} {\bibfnamefont {F.}~\bibnamefont {He\ss{}e}},
  \bibinfo {author} {\bibfnamefont {R.}~\bibnamefont {Lange}},\ and\ \bibinfo
  {author} {\bibfnamefont {S.}~\bibnamefont {Schubert}},\ }\bibfield  {title}
  {\bibinfo {title} {On the dynamics of nonlinear hysteretic systems},\ }in\
  \href@noop {} {\emph {\bibinfo {booktitle} {Vernetzte Wissenschaften,
  Crosslinks in Natural and Social Sciences}}},\ \bibinfo {editor} {edited by\
  \bibinfo {editor} {\bibfnamefont {P.~J.}\ \bibnamefont {Plath}}\ and\
  \bibinfo {editor} {\bibfnamefont {E.-C.}\ \bibnamefont {Ha\ss{}}}}\ (\bibinfo
   {publisher} {Logos Verlag},\ \bibinfo {address} {Berlin},\ \bibinfo {year}
  {2008})\ pp.\ \bibinfo {pages} {271--280}\BibitemShut {NoStop}%
\bibitem [{\citenamefont {Al~Janaideh}\ \emph {et~al.}(2009)\citenamefont
  {Al~Janaideh}, \citenamefont {Rakheja},\ and\ \citenamefont
  {Su}}]{janaideh2009a}%
  \BibitemOpen
  \bibfield  {author} {\bibinfo {author} {\bibfnamefont {M.}~\bibnamefont
  {Al~Janaideh}}, \bibinfo {author} {\bibfnamefont {S.}~\bibnamefont
  {Rakheja}},\ and\ \bibinfo {author} {\bibfnamefont {C.}~\bibnamefont {Su}},\
  }\bibfield  {title} {\bibinfo {title} {A generalized {P}randtl-{I}shlinskii
  model for characterizing the hysteresis and saturation nonlinearities of
  smart actuators},\ }\href@noop {} {\bibfield  {journal} {\bibinfo  {journal}
  {Smart Materials and Structures}\ }\textbf {\bibinfo {volume} {18}},\
  \bibinfo {pages} {045001} (\bibinfo {year} {2009})}\BibitemShut {NoStop}%
\bibitem [{\citenamefont {Shakiba}\ \emph {et~al.}(2018)\citenamefont
  {Shakiba}, \citenamefont {Zakerzadeh},\ and\ \citenamefont
  {A.}}]{Shakiba2018}%
  \BibitemOpen
  \bibfield  {author} {\bibinfo {author} {\bibfnamefont {S.}~\bibnamefont
  {Shakiba}}, \bibinfo {author} {\bibfnamefont {M.~R.}\ \bibnamefont
  {Zakerzadeh}},\ and\ \bibinfo {author} {\bibfnamefont {M.}~\bibnamefont
  {A.}},\ }\bibfield  {title} {\bibinfo {title} {Experimental characterization
  and control of a magnetic shape memory alloy actuator using the modified
  generalized rate-dependent {P}randtl-{I}shlinskii hysteresis model},\
  }\href@noop {} {\bibfield  {journal} {\bibinfo  {journal} {Proceedings of the
  Institution of Mechanical Engineers, Part I: Journal of Systems and Control
  Engineering}\ }\textbf {\bibinfo {volume} {232}},\ \bibinfo {pages} {506}
  (\bibinfo {year} {2018})}\BibitemShut {NoStop}%
\bibitem [{\citenamefont {Liang}\ and\ \citenamefont {Feng}(2020)}]{liang2020}%
  \BibitemOpen
  \bibfield  {author} {\bibinfo {author} {\bibfnamefont {M.}~\bibnamefont
  {Liang}}\ and\ \bibinfo {author} {\bibfnamefont {Y.}~\bibnamefont {Feng}},\
  }\bibfield  {title} {\bibinfo {title} {Modeling of shape memory alloy
  actuated system using a modified rate-dependent {P}randtl-{I}shlinskii
  hysteresis model},\ }in\ \href@noop {} {\emph {\bibinfo {booktitle} {2020 5th
  International Conference on Advanced Robotics and Mechatronics (ICARM)}}}\
  (\bibinfo {year} {2020})\ pp.\ \bibinfo {pages} {113--118}\BibitemShut
  {NoStop}%
\bibitem [{\citenamefont {Yoong}\ \emph {et~al.}(2021)\citenamefont {Yoong},
  \citenamefont {Su},\ and\ \citenamefont {Yeo}}]{Yoong2021}%
  \BibitemOpen
  \bibfield  {author} {\bibinfo {author} {\bibfnamefont {H.~P.}\ \bibnamefont
  {Yoong}}, \bibinfo {author} {\bibfnamefont {C.~Y.}\ \bibnamefont {Su}},\ and\
  \bibinfo {author} {\bibfnamefont {K.~B.}\ \bibnamefont {Yeo}},\ }\bibfield
  {title} {\bibinfo {title} {Stress-dependent generalized
  {P}randtl-{I}shlinskii hysteresis model of a {N}i{T}i wire with superelastic
  behavior},\ }\href@noop {} {\bibfield  {journal} {\bibinfo  {journal}
  {Journal of Intelligent Material Systems and Structures}\ ,\ \bibinfo {pages}
  {1045389X20983888}} (\bibinfo {year} {2021})}\BibitemShut {NoStop}%
\bibitem [{\citenamefont {Chayratsami}\ and\ \citenamefont
  {Plett}(2020)}]{Chayratsami2020}%
  \BibitemOpen
  \bibfield  {author} {\bibinfo {author} {\bibfnamefont {P.}~\bibnamefont
  {Chayratsami}}\ and\ \bibinfo {author} {\bibfnamefont {G.~L.}\ \bibnamefont
  {Plett}},\ }\bibfield  {title} {\bibinfo {title} {Hysteresis modeling of
  lithium-silicon half cells using {P}randtl-{I}shlinskii model},\ }in\
  \href@noop {} {\emph {\bibinfo {booktitle} {2020 IEEE 16th International
  Conference on Control Automation (ICCA)}}}\ (\bibinfo {year} {2020})\ pp.\
  \bibinfo {pages} {1578--1583}\BibitemShut {NoStop}%
\bibitem [{\citenamefont {Barkhausen}(1919)}]{barkhausen1919}%
  \BibitemOpen
  \bibfield  {author} {\bibinfo {author} {\bibfnamefont {H.}~\bibnamefont
  {Barkhausen}},\ }\bibfield  {title} {\bibinfo {title} {{Z}wei mit {H}ilfe der
  neuen {V}erst{\"a}rker entdeckte {E}rscheinungen},\ }\href@noop {} {\bibfield
   {journal} {\bibinfo  {journal} {Phys. Z}\ }\textbf {\bibinfo {volume}
  {20}},\ \bibinfo {pages} {401} (\bibinfo {year} {1919})}\BibitemShut
  {NoStop}%
\bibitem [{\citenamefont {Falk}\ and\ \citenamefont {Langer}(1998)}]{Falk1998}%
  \BibitemOpen
  \bibfield  {author} {\bibinfo {author} {\bibfnamefont {M.~L.}\ \bibnamefont
  {Falk}}\ and\ \bibinfo {author} {\bibfnamefont {J.~S.}\ \bibnamefont
  {Langer}},\ }\bibfield  {title} {\bibinfo {title} {Dynamics of viscoplastic
  deformation in amorphous solids},\ }\href@noop {} {\bibfield  {journal}
  {\bibinfo  {journal} {Phys. Rev. E}\ }\textbf {\bibinfo {volume} {57}},\
  \bibinfo {pages} {7192} (\bibinfo {year} {1998})}\BibitemShut {NoStop}%
\bibitem [{\citenamefont {Spinu}\ \emph {et~al.}(2004)\citenamefont {Spinu},
  \citenamefont {Stancu}, \citenamefont {Radu}, \citenamefont {Li},\ and\
  \citenamefont {Wiley}}]{Spinu2004}%
  \BibitemOpen
  \bibfield  {author} {\bibinfo {author} {\bibfnamefont {L.}~\bibnamefont
  {Spinu}}, \bibinfo {author} {\bibfnamefont {A.}~\bibnamefont {Stancu}},
  \bibinfo {author} {\bibfnamefont {C.}~\bibnamefont {Radu}}, \bibinfo {author}
  {\bibfnamefont {F.}~\bibnamefont {Li}},\ and\ \bibinfo {author}
  {\bibfnamefont {J.~B.}\ \bibnamefont {Wiley}},\ }\bibfield  {title} {\bibinfo
  {title} {Method for magnetic characterization of nanowire structures},\
  }\href@noop {} {\bibfield  {journal} {\bibinfo  {journal} {IEEE Transactions
  on Magnetics}\ }\textbf {\bibinfo {volume} {40}},\ \bibinfo {pages} {2116}
  (\bibinfo {year} {2004})}\BibitemShut {NoStop}%
\bibitem [{\citenamefont {Rotaru}\ \emph {et~al.}(2011)\citenamefont {Rotaru},
  \citenamefont {Lim}, \citenamefont {Lenormand}, \citenamefont {Diaconu},
  \citenamefont {Wiley}, \citenamefont {Postolache}, \citenamefont {Stancu},\
  and\ \citenamefont {Spinu}}]{Rotaru2011}%
  \BibitemOpen
  \bibfield  {author} {\bibinfo {author} {\bibfnamefont {A.}~\bibnamefont
  {Rotaru}}, \bibinfo {author} {\bibfnamefont {J.}~\bibnamefont {Lim}},
  \bibinfo {author} {\bibfnamefont {D.}~\bibnamefont {Lenormand}}, \bibinfo
  {author} {\bibfnamefont {A.}~\bibnamefont {Diaconu}}, \bibinfo {author}
  {\bibfnamefont {J.~B.}\ \bibnamefont {Wiley}}, \bibinfo {author}
  {\bibfnamefont {P.}~\bibnamefont {Postolache}}, \bibinfo {author}
  {\bibfnamefont {A.}~\bibnamefont {Stancu}},\ and\ \bibinfo {author}
  {\bibfnamefont {L.}~\bibnamefont {Spinu}},\ }\bibfield  {title} {\bibinfo
  {title} {Interactions and reversal-field memory in complex magnetic nanowire
  arrays},\ }\href@noop {} {\bibfield  {journal} {\bibinfo  {journal} {Phys.
  Rev. B}\ }\textbf {\bibinfo {volume} {84}},\ \bibinfo {pages} {134431}
  (\bibinfo {year} {2011})}\BibitemShut {NoStop}%
\bibitem [{\citenamefont {Da~Col}\ \emph {et~al.}(2011)\citenamefont {Da~Col},
  \citenamefont {Darques}, \citenamefont {Fruchart},\ and\ \citenamefont
  {Cagnon}}]{DaCol2011}%
  \BibitemOpen
  \bibfield  {author} {\bibinfo {author} {\bibfnamefont {S.}~\bibnamefont
  {Da~Col}}, \bibinfo {author} {\bibfnamefont {M.}~\bibnamefont {Darques}},
  \bibinfo {author} {\bibfnamefont {O.}~\bibnamefont {Fruchart}},\ and\
  \bibinfo {author} {\bibfnamefont {L.}~\bibnamefont {Cagnon}},\ }\bibfield
  {title} {\bibinfo {title} {Reduction of magnetostatic interactions in
  self-organized arrays of nickel nanowires using atomic layer deposition},\
  }\href@noop {} {\bibfield  {journal} {\bibinfo  {journal} {Applied Physics
  Letters}\ }\textbf {\bibinfo {volume} {98}},\ \bibinfo {pages} {112501}
  (\bibinfo {year} {2011})}\BibitemShut {NoStop}%
\bibitem [{\citenamefont {Prandtl}(1928)}]{Prandtl1928}%
  \BibitemOpen
  \bibfield  {author} {\bibinfo {author} {\bibfnamefont {L.}~\bibnamefont
  {Prandtl}},\ }\bibfield  {title} {\bibinfo {title} {Ein {G}edankenmodell zur
  kinetischen {T}heorie der festen {K}örper},\ }\href@noop {} {\bibfield
  {journal} {\bibinfo  {journal} {Z. Ang. Math. Mech.}\ }\textbf {\bibinfo
  {volume} {8}},\ \bibinfo {pages} {85} (\bibinfo {year} {1928})}\BibitemShut
  {NoStop}%
\end{thebibliography}%

\end{document}